\documentclass{aa}  
\pdfoutput=1
\usepackage{graphicx}
\usepackage{txfonts}
\usepackage{natbib}
\usepackage{twoopt}
\usepackage[breaklinks=true]{hyperref} 
\bibpunct{(}{)}{;}{a}{}{,}             

\usepackage{multirow}
\usepackage{caption}
\usepackage{siunitx}
\usepackage{gensymb}

\usepackage{threeparttablex}

\usepackage{subfig}

\hypersetup{bookmarks,
    colorlinks=true,
    citecolor=blue,
    linkcolor=blue,
    urlcolor=blue,
    }

\begin{document}

   \title{Theoretical studies of carbon isotopic fractionation in reactions of C with C$_{2}$: dynamics, kinetics, and isotopologue equilibria}


   \author{C. M. R.~Rocha and H. Linnartz
          }

   \institute{\centering Laboratory for Astrophysics, Leiden Observatory, Leiden University, P.O. Box 9513, NL-2300 RA Leiden, The Netherlands\\
              \email{romerorocha@strw.leidenuniv.nl}
             }

   \date{Received \today}

 
  \abstract
   {Our current understanding of interstellar carbon fractionation 
   hinges on the interpretation of astrochemical kinetic models. Yet, the various reactions included carry large uncertainties in their (estimated) 
   rate coefficients, notably those involving C with C$_{2}$. 
   }
   {We aim to supply theoretical thermal rate coefficients as a function of the temperature for the gas-phase isotope-exchange reactions $\mathrm{^{13}C}\!+\!\mathrm{^{12}C_{2}}(X^{1}\Sigma_{g}^{+},a^{3}\Pi_{u}){\rightleftharpoons}\mathrm{^{13}C^{12}C}(X^{1}\Sigma_{g}^{+},a^{3}\Pi_{u})\!+\!\mathrm{^{12}C}$ and $\mathrm{^{13}C}\!+\!\mathrm{^{13}C^{12}C}(X^{1}\Sigma_{g}^{+},a^{3}\Pi_{u}){\rightleftharpoons}\mathrm{^{13}C_{2}}(X^{1}\Sigma_{g}^{+},a^{3}\Pi_{u})\!+\!\mathrm{^{12}C}$.}
   {By relying on the large masses of the atoms involved, we employ a variation of the {quasi}-classical trajectory method, with the previously obtained (mass-independent) potential energy surfaces of $\mathrm{C_{3}}$ dictating the 
   forces between the colliding partners.}
   {The calculated rate coefficients within the range of $25\!\leq\!T/\si{\kelvin}\!\leq\!500$ show a positive temperature dependence  
   and are markedly different from previous theoretical estimates. While the forward reactions are fast and inherently exothermic owing to the lower zero-point energy content of the products, the reverse processes have temperature thresholds. 
   For each reaction considered, analytic three-parameter Arrhenius-Kooij formulas are provided that readily interpolate {and} extrapolate the associated forward and backward rates. These forms can further be introduced in 
   astrochemical networks. Apart from the proper kinetic attributes, we also provide equilibrium constants for these processes, confirming their prominence in the overall C fractionation chemistry. In this respect, the $^{13}$C+$\mathrm{^{12}C_{2}}(X^{1}\Sigma_{g}^{+})$ and $^{13}$C+$\mathrm{^{12}C_{2}}(a^{3}\Pi_{u})$ reactions are found to be particularly conspicuous, notably at the typical temperatures of dense molecular clouds. {For these reactions and considering both equilibrium and time-dependent chemistry, theoretical $^{12}$C/$^{13}$C ratios as a function of the gas kinetic temperature are also derived and shown to be consistent with available model chemistry and observational data on $\mathrm{C_{2}}$.}}
   {}


   \maketitle
%

\section{Introduction}\label{sec:intro}

Observations of isotopic abundance ratios in interstellar 
molecules provide an avenue 
for tracking Galactic chemical evolution, from 
stellar nucleosynthesis to dense cloud formation and processing of the ejected material to new stars and planetary systems created therefrom~\citep{WIL99:143}. 
For example, the seemingly incompatible elemental [$^{12}$C/$^{13}$C] ratios 
found in the local interstellar medium (ISM; $\sim\!68$ as inferred 
from CN~\citep{MIL005:1126}, CO~\citep{LAN92:193}, H$_{2}$CO~\citep{LAN92:193} and CH$^{+}$~\citep{WIL99:143}) and in the Solar System ($\sim\!89$) might be 
indicative of $^{13}$C enrichment of the ISM by asymptotic giant branch (AGB) stars since the formation of the Sun~\citep{MIL005:1126}. 

Apart from the intrinsic variations with galactocentric distance and time~\citep{WIL99:143,MIL005:1126,LAN92:193}, isotopic 
abundance ratios as measured in molecules are also important tracers of local  environment effects. Interstellar species 
often show relative abundances of particular isotopologs that may significantly differ from those inherent in the gas owing to peculiarities in their chemistry~\citep{FUR011:38}. 
In cold dense cloud cores, with typical temperatures ($T$) of $\!\sim\!10\,\si{\kelvin}$ and 
visual extinctions ($A_{V}$) of $\!\sim\!10\,\mathrm{mag}$, 
this so-called isotopic fractionation~\citep{LAN84:581,TER000:563,FUR011:38,LIS012:55,ROU015:A99,FUR018:105,
LOI018:2447,LOI019:5777,COL020,LOI020} has long been recognized and mainly attributed to gas-phase 
isotope-exchange reactions~\citep{DAL76:573,WAT76:L165}.    
Given the very low collision energies in dense clouds, 
it becomes clear that the most efficient fractionation pathways therein must involve exothermic 
reactions for which the salient features of the potential 
energy surfaces~{\citep[PESs;][]{ROC020}} are basins rather than barriers~\citep{HEN89:1673}. Indeed, chemical fractionation via barrierless ion--molecule or neutral--neutral reactions is mostly driven by the small zero-point energy (ZPE) differences between reactants and products of isotopically distinct species~\citep{MLA014:A144,MLA017:A22}; the role of isotope-selective gas--grain interactions and photodissociation in also altering fractionation ratios is discussed elsewhere~\citep[\emph{e.g.},][]{FUR011:38,FUR018:105,LOI018:2447,VIS009:323}.

With regard to carbon isotopic fractionation, 
~\citet{WAT76:L165} first pointed out the relevance of the reaction 
\begin{equation}\label{eq:reacbase}
\mathrm{^{13}C^{+}}+\mathrm{^{12}CO}\underset{k_{\text{-}1}}{\stackrel{k_{1}}{\rightleftharpoons}} \mathrm{^{13}CO}+\mathrm{^{12}C^{+}}+\Delta E_{\mathrm{ZPE}}^{(1)}
,\end{equation}
which is particularly efficient at low $T$; 
$k_{1}/k_{\text{-}1}\!\approx\!33$ at 10\,\si{\kelvin} and $\Delta E_{\mathrm{ZPE}}^{(1)}$, the ZPE difference among 
$\mathrm{^{12}CO}$ and $\mathrm{^{13}CO}$, is $\approx\!35\,\si{\kelvin}$~\citep{WAT76:L165,SMI80:424,LAN84:581}. As first noted by
~\citet{LAN84:581} 
reaction~(\ref{eq:reacbase}), on one hand, enhances the amount of $^{13}$C locked up in CO (and in species directly formed from it), and on the
other hand makes  $\mathrm{^{13}C^{+}}$ less available to react with other C-bearing species, decreasing their $\mathrm{^{13}C}$ content. 
Because CO is by far the largest repository of gas-phase carbon (at least in oxygen-rich dense clouds~\citep{LAN84:581}), the above scenario 
led to the suggestion that $^{12}$C/$^{13}$C values as measured from CO serve as a lower limit to the ``true'' elemental [$^{12}$C/$^{13}$C] ratio gradient throughout the Galaxy, while those inferred from other species like H$_{2}$CO reflect an upper range~\citep{WIL99:143,LAN92:193}.

Ever since the postulation of reaction~(\ref{eq:reacbase}) as the main 
C fractionation route in strongly
shielded regions~\citep{WAT76:L165}, a notable contrast {has emerged}  
between the above general
predictions (by chemistry models) of the strong $^{13}$C depletion in C-containing molecules~\citep{LAN84:581} 
and the general absence of this observable effect in surveys conducted, for example, in abundant species such as  
$\mathrm{CS}$~\citep{LIS012:55}, 
$\mathrm{CN}$~\citep{MIL005:1126},
$\mathrm{C_{2}}$~\citep{HAM019:143}, 
$\mathrm{CCS}$~\citep{SAK007:1174},
$\mathrm{HNC}$~\citep{LIS012:55},
$\mathrm{C_{3}}$~\citep{GIE020:A120}, 
and $\mathrm{HC_{3}N}$~\citep{TAK98:1156} that are not formed directly from CO and whose $^{12}$C/$^{13}$C ratios thus inferred are in agreement with (or even lower than) the gas elemental values. 
Such a conflict therefore opened up new avenues for the possibility of an overall $^{13}$C enrichment in species other than CO, and led to the proposition of alternative isotope-exchange reactions (\emph{e.g.}, $^{13}$C$^{(+)}$+CN~\citep{LAN92:193,ROU015:A99}, $^{13}$CO+HCO$^{+}$~\citep{SMI80:424,MLA017:A22},  $^{13}$C+C$_{2}$~\citep{ROU015:A99}, and $^{13}$C+C$_{3}$~\citep{GIE020:A120,COL020,LOI020}) 
and novel formation pathways~\citep{TAK98:1156,SAK007:1174,FUR011:38} deemed to contribute to the $^{13}$C fractionation chemistry. Despite 
previous assessments~\citep{WOO009:1360,FUR011:38,ROU015:A99,COL020,LOI020}, 
validation of {this hypothesis is often} hindered by a lack of accurate experimental and/or theoretical rate coefficients for some of these reactions~\citep{FUR011:38,WOO009:1360}.

In this work, we provide such values for the gas-phase reactions
\begin{equation}\label{eq:reac1}
\mathrm{^{13}C}+\mathrm{^{12}C_{2}}(X^{1}\Sigma_{g}^{+})\underset{k_{\text{-}2}}{\stackrel{k_{2}}{\rightleftharpoons}}\mathrm{^{13}C^{12}C}(X^{1}\Sigma_{g}^{+})+\mathrm{^{12}C}+\Delta E_{\mathrm{ZPE}}^{(2)},
\end{equation}
\begin{equation}\label{eq:reac2}
\mathrm{^{13}C}+\mathrm{^{13}C^{12}C}(X^{1}\Sigma_{g}^{+})\underset{k_{\text{-}3}}{\stackrel{k_{3}}{\rightleftharpoons}}\mathrm{^{13}C_{2}}(X^{1}\Sigma_{g}^{+})+\mathrm{^{12}C}+\Delta E_{\mathrm{ZPE}}^{(3)},
\end{equation} 
\begin{equation}\label{eq:reac3}
\mathrm{^{13}C}+\mathrm{^{12}C_{2}}(a^{3}\Pi_{u})\underset{k_{\text{-}4}}{\stackrel{k_{4}}{\rightleftharpoons}}\mathrm{^{13}C^{12}C}(a^{3}\Pi_{u})+\mathrm{^{12}C}+\Delta E_{\mathrm{ZPE}}^{(4)},
\end{equation}
and
\begin{equation}\label{eq:reac4}
\mathrm{^{13}C}+\mathrm{^{13}C^{12}C}(a^{3}\Pi_{u})\underset{k_{\text{-}5}}{\stackrel{k_{5}}{\rightleftharpoons}}\mathrm{^{13}C_{2}}(a^{3}\Pi_{u})+\mathrm{^{12}C}+\Delta E_{\mathrm{ZPE}}^{(5)},
\end{equation}
by means of a theoretical approach (see below). 
The motivation here is primarily grounded in the 
prevalence of $\mathrm{C_{2}}$, the smallest pure carbon cluster, throughout 
the ISM; it has been detected (via its Phillips ($A\,^{1}\Pi_{u}$--$X\,^{1}\Sigma_{g}^{+}$) and Swan ($d\,^{3}\Pi_{g}$--$a\,^{3}\Pi_{u}$) bands) in a myriad of astronomical sources~\citep{BAB019:38}, including 
diffuse~\citep{SOU77:L49,SNO006:367}, translucent~\citep{HAM019:143}, and 
dense molecular clouds~\citep{HOB83:L95} and is known to be the primary reservoir of gas-phase   carbon in oxygen-poor regions~\citep{SOU77:L49}. 
Besides being key for probing the physical conditions of interstellar clouds~\citep{SNO006:367}, 
$\mathrm{C_{2}}$, together with C$^{(+)}$, is thought to be the fundamental 
building block in the formation chemistry 
of larger hydrogen-deficient C-bearing species~\citep{EHR000:427,KAI002:1309,GU006:245}, and therefore plays an active role in their $^{13}$C enrichment. From a top-down perspective, $\mathrm{C_{2}}$ {radicals} are also important units arising 
from the (photo)fragmentation of polycyclic aromatic hydrocarbons (PAHs) and 
fullerenes. {In a number of experimental studies it was shown that PAHs, once fully dehydrogenated, fragment through sequential C$_{2}$-losses (see, \emph{e.g.},~\citet{ZHE014:L30})}. This is fully consistent with the general picture that some of the diffuse interstellar band (DIB) carriers, notably those responsible for the so-called C$_{2}$ DIBs~\citep{THO003:339,ELY018:A143}, might {be related to PAH cations and their derivatives upon photoprocessing}. 

As for the calculation of both forward and reverse rate coefficients of reactions~(\ref{eq:reac1})-(\ref{eq:reac4}), we herein employ the {quasi}-classical trajectory (QCT) method~\citep{TRU79:505,PES99:171}, with the previously obtained (nuclear-mass-independent) global PESs of $\mathrm{C_{3}}(^{3}A')$~\citep{ROC019:8154} and $\mathrm{C_{3}}(^{1}A')$~\citep{ROC018:36} dictating the interactions 
between the involved nuclei (see Section~\ref{sec:calcdet}). 
From the calculated rate coefficients as a function of $T$, 
equilibrium constants for these processes are also provided 
and their possible impact on the overall C isotopic fractionation chemistry 
is  
briefly discussed. 

\section{Methods}\label{sec:calcdet}
\subsection{Potential  energy surfaces}\label{subsec:pess}
\begin{figure}
\centering
\includegraphics[angle=0,width=1\linewidth]{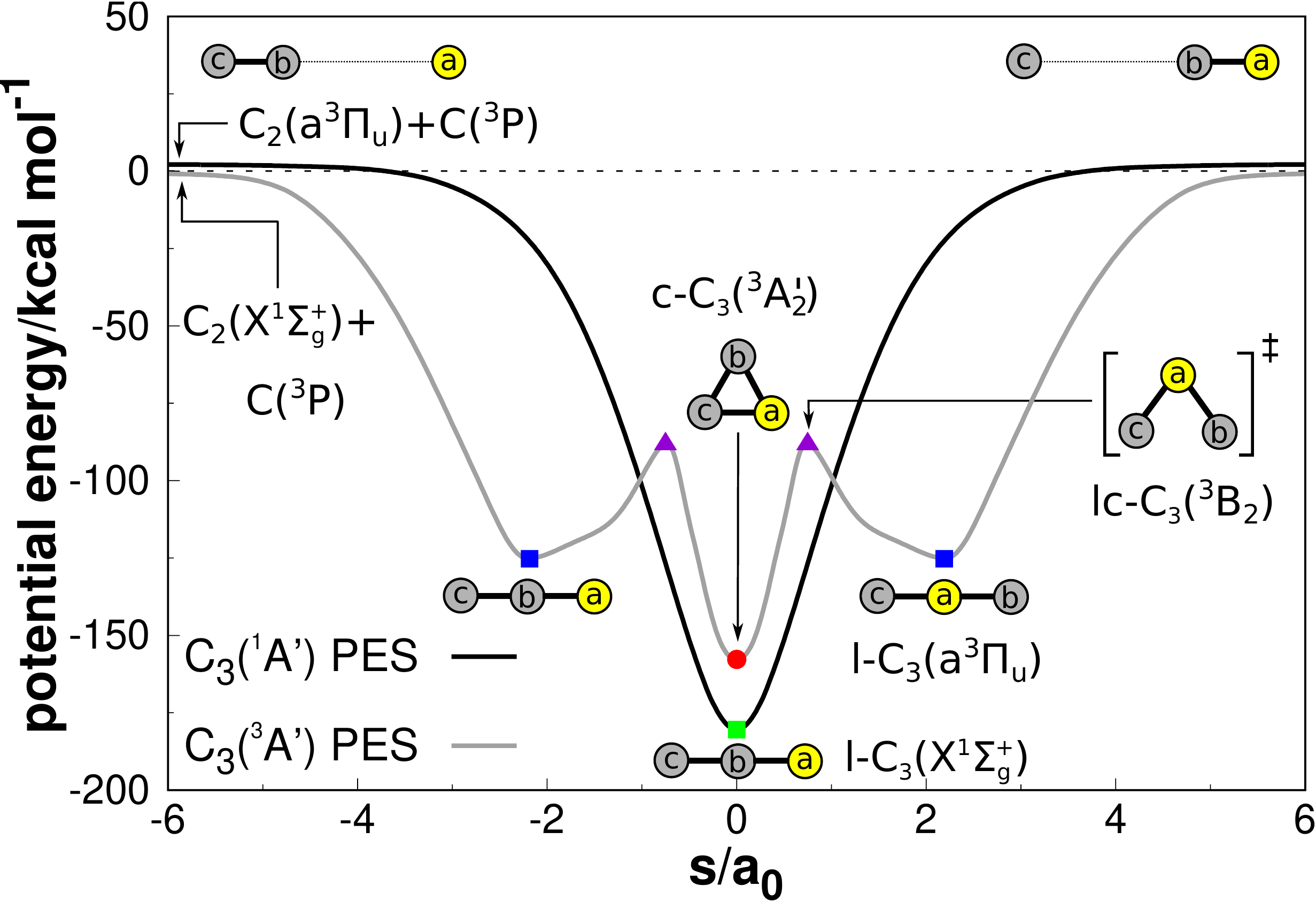}
\caption{\footnotesize One-dimensional cuts of the nuclear-mass-independent PESs of 
$\mathrm{C_{3}}$ 
along the minimum-energy paths connecting reactants and products via $\mathrm{C_{3}}$ intermediates. The zero
of energy is set relative to the infinitely separated C+C$_{2}(X\,^{1}\Sigma_{g}^{+})$ fragments. The reacting carbon atom is shown in yellow.} 
\label{fig:mep}
\end{figure}
\begin{figure}
\captionsetup[subfigure]{labelformat=empty}
\centering
\subfloat{{\includegraphics[width=1\linewidth]{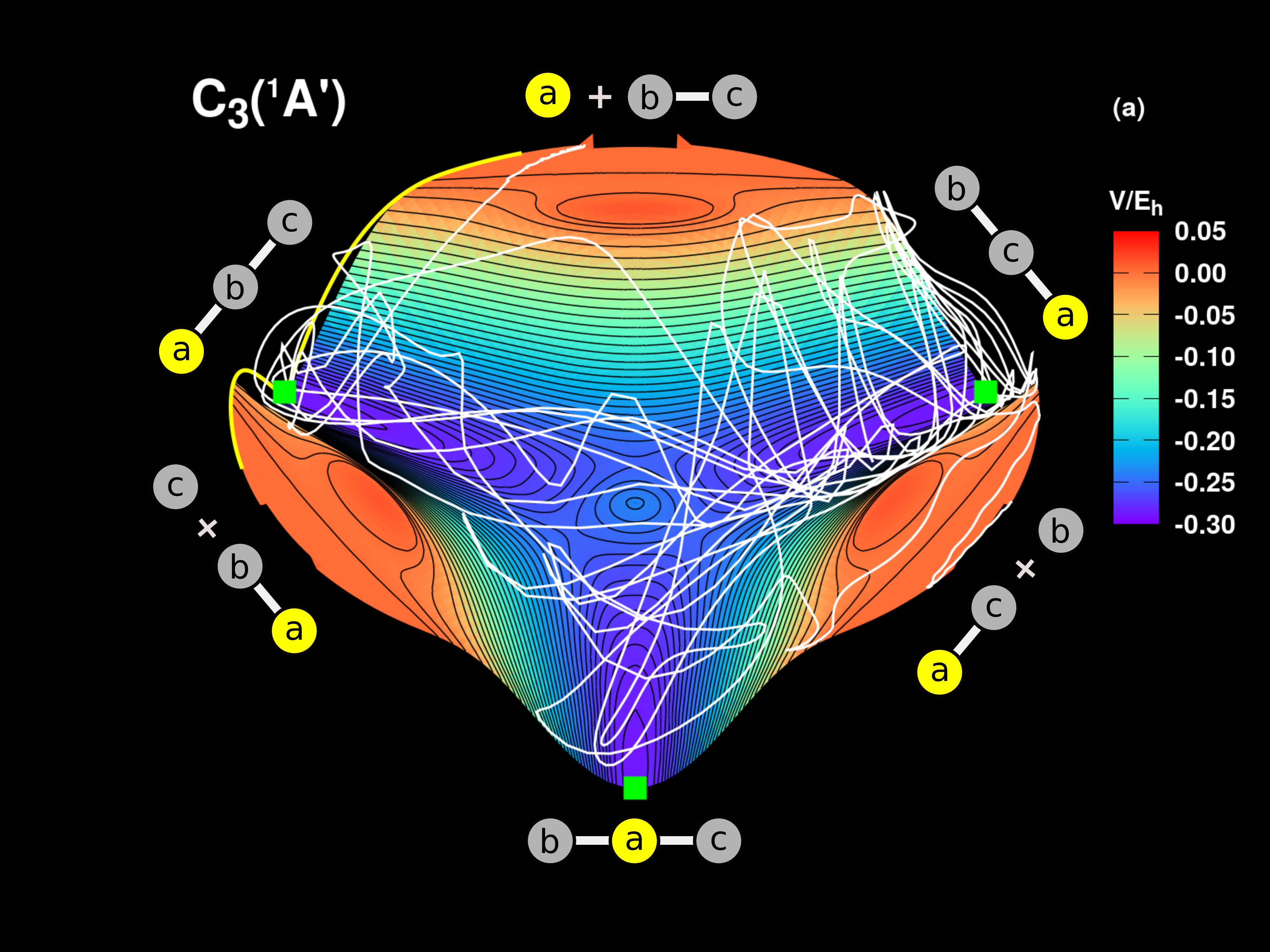}}}
\vfill 
\subfloat{{\includegraphics[width=1\linewidth]{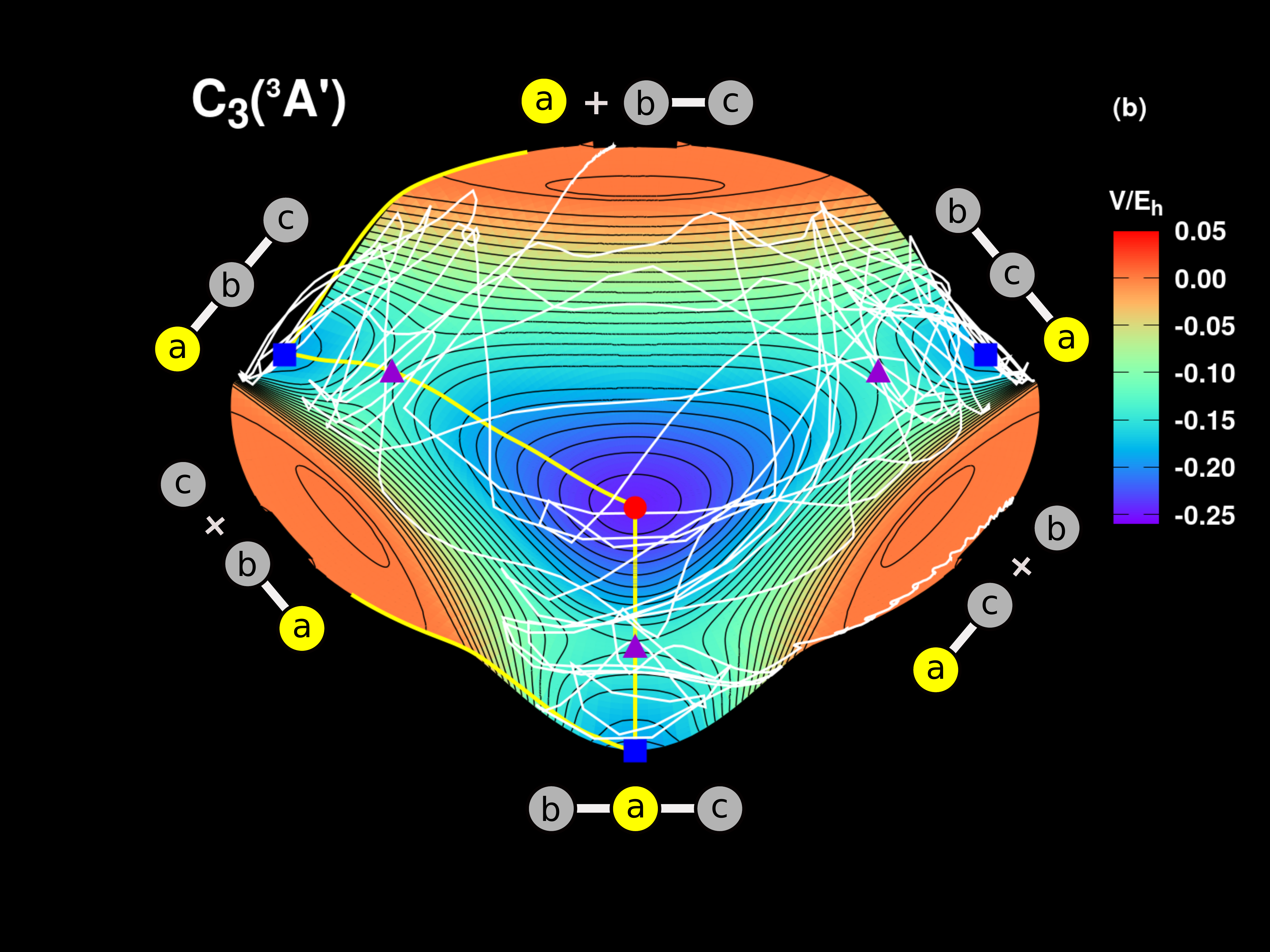}}}
\caption{\footnotesize Relaxed 3$D$ plots in hyperspherical 
coordinates~\citep{VAR87:455} of the nuclear-mass-independent PESs of ground-state~(a)~$\mathrm{C_{3}}(^{1}A')$~and~(b)~$\mathrm{C_{3}}(^{3}A')$. The zero of energy is set relative to the infinitely separated C+C$_{2}$ fragments. 
Stationary points and minimum energy paths (solid yellow lines) as in Figure~\ref{fig:mep}. Solid white lines show the time evolution (in coordinate space) of sample reactive trajectories for the processes~(a)$\mathrm{^{13}C}(^{3}P)\!+\!\mathrm{^{12}C_{2}}(a^{3}\Pi_{u})\!\rightarrow\!\mathrm{^{13}C^{12}C}(a^{3}\Pi_{u})\!+\!\mathrm{^{12}C}(^{3}P)$~and~(b)~$\mathrm{^{13}C}(^{3}P)\!+\!\mathrm{^{12}C_{2}}(X^{1}\Sigma_{g}^{+})\!\rightarrow\!\mathrm{^{13}C^{12}C}(X^{1}\Sigma_{g}^{+})\!+\mathrm{^{12}C}(^{3}P)$. An isotopically distinct carbon atom is schematically represented in yellow. 
} 
\label{fig:pess}
\end{figure}
\begin{table*}
\begin{footnotesize}
\begin{center}
\caption{\footnotesize Spectroscopic properties (in $\rm cm^{-1}$) of the 
reactant and product diatomics (of reactions~(\ref{eq:reac1})-(\ref{eq:reac4})) correlating with the global singlet and triplet PESs of $\mathrm{C_{3}}$.}
\label{tab:diat}
\begin{threeparttable}
\begin{tabular}{ll
S[table-align-text-post=false,table-format=3.1]
S[table-align-text-post=false,table-format=1.3]
S[table-align-text-post=false,table-format=4.1]
S[table-align-text-post=false,table-format=2.4]
S[table-align-text-post=false,table-format=2.4]
S[table-align-text-post=false,table-format=1.4]
S[table-align-text-post=false,table-format=1.4]
S[table-align-text-post=false,table-format=3.1]}
\hline\hline \\[-1.8ex]
 & Source & {$T_{e}$\tnote{a}} & {$R_{e}$} & $\omega_{e}$ & {$\omega_{e}x_{e}$} & {$\omega_{e}y_{e}$} & {$B_{e}$} & {$\alpha_{e}$} & {$E_{\mathrm{ZPE}}$} \\[0.5ex] 
\hline \\[-2.25ex]
 \multirow{2}{*}{\centering $\mathrm{^{12}C_{2}}(X\,^{1}\Sigma_{g}^{+})$}        & $\mathrm{C_{3}}(^{3}A')$ PES  &    0.0  & 2.348 & 1855.5 & 13.5624 & -0.1655 & 1.8203 & 0.0214 & 924.1 \\
                                                                                                                                              & exp.\tnote{b}                                 &    0.0  & 2.348 & 1855.0 & 13.5701 & -0.1275 & 1.8200 & 0.0179 & 924.1 \\
 \\ 
  \multirow{2}{*}{\centering $\mathrm{^{13}C^{12}C}(X\,^{1}\Sigma_{g}^{+})$}  &  $\mathrm{C_{3}}(^{3}A')$ PES  &    0.0  & 2.348 & 1819.5 & 13.1000 & -0.1480 & 1.7503 &  0.0210 & 906.2 \\ 
                                                                                                                                              & exp.\tnote{b,c}                                  &    0.0  & 2.348 &  1818.9  & 13.0466 &  -0.1202             &  1.7498           &  0.0169  &  906.2 \\
 \\
   \multirow{2}{*}{\centering $\mathrm{^{13}C_{2}}(X\,^{1}\Sigma_{g}^{+})$}  &  $\mathrm{C_{3}}(^{3}A')$ PES  &    0.0  & 2.348 & 1782.7 & 12.5845 & -0.1400 & 1.6803 & 0.0342  & 887.9 \\ 
                                                                                                                                              & exp.\tnote{b}                                  &    0.0  & 2.348 &  1781.8  & 12.3560 &  -0.1466             &      1.6796       &  0.0157  & 887.8 \\
 \\
\hline 
 \\
 \multirow{2}{*}{\centering $\mathrm{^{12}C_{2}}(a^{3}\Pi_{u})$}                           &  $\mathrm{C_{3}}(^{1}A')$ PES   & 715.3  & 2.479 & 1641.3 & 11.5723 & -0.0047  & 1.6324 &  0.0177 & 819.5 \\ 
                                                                                                                                              & exp.\tnote{b}                                   & 716.2  & 2.479 & 1641.3 & 11.6595 &                & 1.6323 &  0.0166 & 819.4 \\
\\
 \multirow{2}{*}{\centering $\mathrm{^{13}C^{12}C}(a^{3}\Pi_{u})$}                      &  $\mathrm{C_{3}}(^{1}A')$ PES    & 715.3  & 2.479 & 1609.5 & 11.1471 & -0.0037  & 1.5696 &  0.0167  & 803.6 \\ 
                                                                                                                                              & exp.\tnote{b}                                     & 716.2  & 2.479 & 1609.4 & 11.2103 &               & 1.5693 &  0.0156  & 803.4 \\
\\
 \multirow{2}{*}{\centering $\mathrm{^{13}C_{2}}(a^{3}\Pi_{u})$}                      &  $\mathrm{C_{3}}(^{1}A')$ PES    & 715.3  & 2.479 & 1576.9 & 10.7539 & -0.0012  & 1.5068 &  0.0293  & 787.3 \\ 
                                                                                                                                              & exp.\tnote{b}                                     & 716.2  & 2.479 &  &  &               & 1.4993 &    & \\                                                                                                                                              
\hline\hline
\end{tabular}
\begin{tablenotes}[flushleft]
\item[a]{{\footnotesize Energies given 
with respect to the corresponding ground electronic states of each isotopologue}.}
    \item[b]{{\footnotesize Data from \citet{AMI83:329}, \citet{BRO013:11}, 
    \citet{RAM014:5}, and~\citet{WAN015:064317}}.}
    \item[c]{{\footnotesize Experimental spectroscopic constants calculated 
    from $\mathrm{^{12}C_{2}(X^{1}\Sigma_{g}^{+})}$ data and isotopic relationships~{\citep[see, \emph{e.g.},][]{RAM014:5}}.}}    
\end{tablenotes}
\end{threeparttable}
\end{center}
\end{footnotesize}
\end{table*}
\begin{table*}
\begin{footnotesize}
\begin{center}
\caption{\footnotesize Structural parameters (bond distances $R_{e}$ in \si{\bohr} and angle $\alpha$ in degs), 
harmonic ($\omega_{i}$), fundamental ($\nu_{i}$) frequencies and zero-point energies (in $\rm cm^{-1}$)  
of the $\mathrm{^{13}C}$ singly and doubly substituted $\mathrm{C_{3}}$ minima of the PESs spanned by long-lived trajectories.}
\label{tab:triat}
\begin{threeparttable}
\begin{tabular}{ll
S[table-align-text-post=false,table-format=1.3]
S[table-align-text-post=false,table-format=3.1]
S[table-align-text-post=false,table-format=4.1]
S[table-align-text-post=false,table-format=4.1]
S[table-align-text-post=false,table-format=4.1]
S[table-align-text-post=false,table-format=4.1]
S[table-align-text-post=false,table-format=4.1]
S[table-align-text-post=false,table-format=4.1]
S[table-align-text-post=false,table-format=4.1]}
\hline\hline \\[-1.8ex]
 & Source & {$R_{e}$\tnote{a}} & {$\alpha$} & $\omega_{1}$\tnote{b} & {$\omega_{2}$\tnote{b}} & {$\omega_{3}$\tnote{b}} & {$\nu_{1}$\tnote{b}} & {$\nu_{2}$\tnote{b}} & {$\nu_{3}$\tnote{b}} & {$E_{\mathrm{ZPE}}$} \\[0.5ex] 
\hline \\[-2.25ex]
 \multirow{2}{*}{\centering $\ell$-$\mathrm{^{13}C^{12}C^{12}C}(X\,^{1}\Sigma_{g}^{+})$}        & $\mathrm{C_{3}}(^{1}A')$ PES  & 2.445  & 180.0 & 1182.8 & 42.5 & 2088.0 & 1201.9 &  63.2 &  2027.7 &  1686.5 \\
                                                                                                                                                                & exp.\tnote{c}                                 &  2.445  &           &             &         &             &   & 63.1 &  2027.1 &  \\
\\
 \multirow{2}{*}{\centering $\ell$-$\mathrm{^{12}C^{13}C^{12}C}(X\,^{1}\Sigma_{g}^{+})$}        & $\mathrm{C_{3}}(^{1}A')$ PES  & 2.445  & 180.0 & 1206.7 & 41.7 & 2046.7 &  1212.1 &  62.2 & 2008.5 &  1681.0 \\
                                                                                                                                                                & exp.\tnote{c}                                 &  2.445  &  180.0  &            &         &             &  & 61.1 &  &  \\
\\ 
 \multirow{2}{*}{\centering $\ell$-$\mathrm{^{13}C^{12}C^{13}C}(X\,^{1}\Sigma_{g}^{+})$}        & $\mathrm{C_{3}}(^{1}A')$ PES  & 2.445  & 180.0 & 1159.4 & 42.2 & 2074.2  & 1189.2 & 61.9 & 1995.2 & 1662.3 \\
                                                                                                                                                                & exp.\tnote{c}                                 &  2.445  &  180.0  &            &         &             &  & 62.9 &  &  \\
\\ 
\multirow{2}{*}{\centering $\ell$-$\mathrm{^{13}C^{13}C^{12}C}(X\,^{1}\Sigma_{g}^{+})$}        & $\mathrm{C_{3}}(^{1}A')$ PES  & 2.445  & 180.0 & 1182.8 & 41.4 & 2033.1 & 1199.2 & 60.8 & 1976.4 & 1656.9 \\
                                                                                                                                                                & exp.\tnote{c}                                 &  2.445  &  180.0  &            &         &             &  & 60.7 &  &  \\
\\
\hline
\\
 \multirow{1}{*}{\centering $c$-$\mathrm{^{13}C^{12}C^{12}C}(^{3}A'_{2})$}        & $\mathrm{C_{3}}(^{3}A')$ PES  & 2.580  & 60.0 & 1530.5 & 1077.6 & & 1503.8  & 1081.1  & & 1851.6 \\                                                                                                                                                                
\\
 \multirow{1}{*}{\centering $c$-$\mathrm{^{13}C^{12}C^{13}C}(^{3}A'_{2})$}        & $\mathrm{C_{3}}(^{3}A')$ PES  & 2.580  & 60.0 & 1511.0 & 1063.3 & & 1483.6  & 1067.1  & & 1826.8 \\                                                                                                                                                                
\\
 \hline
 \\
  \multirow{1}{*}{\centering $\ell$-$\mathrm{^{13}C^{12}C^{12}C}(\tilde{a}\,^{3}\Pi_{u})$}        & $\mathrm{C_{3}}(^{3}A')$ PES  & 2.466  & 180.0 & 1122.1 & 550.2 & 1331.5 &  1142.5 & 495.7  & 1403.0  &  1782.7  \\
                                                                                                                                                                & exp.\tnote{d}                                 &  2.465  &  180.0 &             &         &             &   & &  &  \\          
 \\
   \multirow{1}{*}{\centering $\ell$-$\mathrm{^{12}C^{13}C^{12}C}(\tilde{a}\,^{3}\Pi_{u})$}        & $\mathrm{C_{3}}(^{3}A')$ PES  & 2.466  & 180.0 & 1145.3 & 539.4 & 1304.5 &  1153.5 & 491.4  &  1388.6 &  1776.4 \\
                                                                                                                                                                & exp.\tnote{d}                                 & 2.465  & 180.0  &             &         &             &   & &  &  \\   
\\
   \multirow{1}{*}{\centering $\ell$-$\mathrm{^{13}C^{12}C^{13}C}(\tilde{a}\,^{3}\Pi_{u})$}        & $\mathrm{C_{3}}(^{3}A')$ PES  & 2.466  & 180.0 & 1100.4 & 547.1 & 1322.1 & 1131.2  & 488.5  & 1379.2  & 1757.1  \\
                                                                                                                                                                & exp.\tnote{d}                                 & 2.465  & 180.0  &             &         &             &   & &  &  \\ 
\\
   \multirow{1}{*}{\centering $\ell$-$\mathrm{^{13}C^{13}C^{12}C}(\tilde{a}\,^{3}\Pi_{u})$}        & $\mathrm{C_{3}}(^{3}A')$ PES  & 2.466  & 180.0 & 1121.8 & 536.2 & 1296.8 & 1141.4 & 484.1 & 1365.8  &  1751.0 \\
                                                                                                                                                                & exp.\tnote{d}                                 & 2.465  & 180.0  &             &         &             &   & &  &  \\                                                                                                                                                                 
\hline\hline
\end{tabular}
\begin{tablenotes}[flushleft]
\item[a]{{\footnotesize $R_{e}\!=\!R_{1}\!=\!R_{2}$}.}
\item[b]{{\footnotesize See \citet{ROC018:36,ROC019:8154} for the definition of the vibrational modes and to assess the corresponding values for the main isotopologues}.}  
\item[c]{{\footnotesize Data from \citet{KRI013:3332} and \citet{BRE016:234302}}.}
\item[d]{{\footnotesize Data from \citet{TOK95:3928}}.}
\end{tablenotes}
\end{threeparttable}
\end{center}
\end{footnotesize}
\end{table*}
\begin{table*}
\begin{footnotesize}
\begin{center}
\caption{\footnotesize Exothermicities (in $\rm cm^{-1}$ unless otherwise stated) of reactions~(\ref{eq:reac1})-(\ref{eq:reac4}) based on the data shown in Table~\ref{tab:diat}.}
\label{tab:zpe}
\begin{threeparttable}
\begin{tabular}{cll
S[table-align-text-post=false,table-format=3.8]
S[table-align-text-post=false,table-format=3.8]
S[table-align-text-post=false,table-format=3.8]
S[table-align-text-post=false,table-format=3.8]}
\hline\hline \\[-1.8ex]
 & & & \multicolumn{4}{c}{\centering Reaction~\#} \\ 
\cline{4-7}\\[-0.2cm]
 & & Source & {(\ref{eq:reac1})} & {(\ref{eq:reac2})} & {(\ref{eq:reac3})} & {(\ref{eq:reac4})} \\
\hline \\[-2.25ex]
  \multirow{3}{*}{\centering $\Delta E_{\mathrm{ZPE}}$\tnote{a,b}} & & this work & 17.9{\,(25.8\,\si{\kelvin})} & 18.3{\,(26.3\,\si{\kelvin})} & 15.9{\,(22.9\,\si{\kelvin})} & 16.3{\,(23.5\,\si{\kelvin})}\\
& & others\tnote{c}                    & 18.0{\,(25.9\,\si{\kelvin})} & 18.3{\,(26.4\,\si{\kelvin})} &                              &                             \\
& & exp.\tnote{d}                      & 17.9{\,(25.8\,\si{\kelvin})} & 18.4{\,(26.5\,\si{\kelvin})} & 16.0{\,(23.0\,\si{\kelvin})} &                             \\
\hline\hline
\end{tabular}
\begin{tablenotes}[flushleft]
\item[a]{{\footnotesize This assumes
that the reactions proceed in the ground-rovibrational states of
both the reactants and products}.}
\item[b]{{\footnotesize The corresponding zero point energies in \si{\kelvin}, $\Delta E_{\mathrm{ZPE}}/k_{B}$, are also given in parenthesis}.}
\item[c]{{\footnotesize Data from \citet{COL020}}.}
\item[d]{{\footnotesize Experimental estimates using the data from Table~\ref{tab:diat}}.}    
\end{tablenotes}
\end{threeparttable}
\end{center}
\end{footnotesize}
\end{table*}
The global adiabatic mass-independent PESs of ground-state $\mathrm{C_{3}}(^{1}A')$  
and $\mathrm{C_{3}}(^{3}A')$ 
used here in the QCT calculations 
are depicted in Figures~\ref{fig:mep} and~\ref{fig:pess}. They were obtained by performing electronic structure calculations for a sufficient number of (fixed) 
nuclear configurations whose energies were then modeled by physically motivated many-body expansion forms~\citep{ROC018:36,ROC019:8154}. To obtain a balanced and accurate description of both valence and long-range features of the potentials, \emph{ab initio} calculations were carried out at the 
multireference configuration interaction [MRCI(+Q)] level of theory~\citep{SZA012:108}, with the final total energies subsequently extrapolated to the complete (one-electron) basis set limit~\citep{VAR018:177} prior 
to the fitting procedure. For the singlet PES, \citet{ROC018:36} improved the spectroscopy near its linear minima [$\ell$-$\mathrm{C_{3}}(X\,^{1}\Sigma_{g}^{+})$] by morphing this global form with an accurate Taylor-series expansion taken from~\citet{SCH016:044307}. In this spirit 
and to partially account for the incompleteness of the $\mathcal{N}$-electron basis and other minor effects, 
both global PESs used in this work have 
their \emph{ab initio} two-body terms  
replaced by the direct-fit, experimentally determined, diatomic curves~\citep{ROC019:8154}. The spectroscopic attributes of the isotopically substituted dissociation channels and of the $\mathrm{C_{3}}$ intermediates spanned by the 
trajectories are shown in Tables~\ref{tab:diat} and~\ref{tab:triat}, respectively. Also listed in Table~\ref{tab:zpe} are the corresponding $\Delta E_{\mathrm{ZPE}}$ values 
of reactions~(\ref{eq:reac1})-(\ref{eq:reac4}) as predicted from the global PESs; their thermodynamic aspects are briefly summarized below. 

As Figure~\ref{fig:mep} shows, the underlying C exchange reactions proceed without activation barriers for collinear atom--diatom approaches. Along $C_{\infty v}$, the shape of the ground-state $\mathrm{C_{3}}(^{1}A')$ PES is characterized by a single, deep potential well; the stabilization 
energy of the $\ell$-$\mathrm{C_{3}}(X\,^{1}\Sigma_{g}^{+})$ complex is $183\,\mathrm{kcal\,mol^{-1}}$ relative to 
the infinitely separated $\mathrm{C}(^{3}P)$+$\mathrm{C_{2}}(a^{3}\Pi_{u})$ fragments. In contrast, the minimum energy path (MEP) for the $\mathrm{C}(^{3}P)$+$\mathrm{C_{2}}(X\,^{1}\Sigma_{g}^{+})$ insertion 
unravels the existence of two such wells; the shallower of the two with a well depth of $-125\,\mathrm{kcal\,mol^{-1}}$ characterizes the 
$\ell$-$\mathrm{C_{3}}(\tilde{a}\,^{3}\Pi_{u})$ local minimum, 
while the deepest at $-158\,\mathrm{kcal\,mol^{-1}}$ defines 
the $c$-$\mathrm{C_{3}}(^{3}A'_{2})$ equilateral triangular global minimum. The access from one basin to the other is granted 
via the $C_{2v}$ transition state (TS) $\ell c$-$\mathrm{C_{3}}(^{3}B_{2})$ with activation energy of $37\,\mathrm{kcal\,mol^{-1}}$ relative to $\ell$-$\mathrm{C_{3}}(\tilde{a}\,^{3}\Pi_{u})$. We note that, due to the permutational nature of the PESs, three symmetry-equivalent and interconnected MEPs exist for rotations by $\pm\,120\degree$ (see Figure~\ref{fig:pess}); this is expected to enhance the efficiency of the isotopic scrambling by long-lived $\mathrm{C_{3}}$ intermediates~\citep{HEN89:1673}. 
However, differently from the collinear insertions, Figure~\ref{fig:pess} unravels the presence of energy barriers along 
perpendicular approaches of 
the fragments; these are $\approx\!9$ and $2\,\mathrm{kcal\,mol^{-1}}$ for the $\mathrm{C_{3}}(^{1}A')$ and $\mathrm{C_{3}}(^{3}A')$ PESs, respectively, and therefore make reactive events arising from 
$C_{2v}$ atom--diatom encounters prohibitive at low $T$. 

\subsection{Quasi-classical trajectory calculations}\label{subsec:qct}
The quasi-classical trajectory (QCT) method employed in this work has been extensively  
described in the literature~\citep{TRU79:505,PES99:171}. Using a locally modified 
version of the VENUS96C code~\citep{VENUS}, batches of $10^{4}$ %
trajectories were run for the ground
adiabatic $^{1}A'$ 
and $^{3}A'$ 
PESs of $\mathrm{C_{3}}$ 
separately; non-adiabatic~\citep{TUL71:562,VOR98:6057,GAL012:22A515} 
and spin-forbidden~\citep{TAC95:16630,GAL013:2292} transitions were not taken into account. 
Cross-sections and rate constants for the envisaged (forward and reverse) isotope-exchange reactions [Eqs.~(\ref{eq:reac1})-(\ref{eq:reac4})] were 
obtained for fixed $T$s by randomly sampling~\citep{PES99:171}  
the orientation of the reactants; atom-diatom 
relative translational energy; the ro-vibrational state of the reactant dicarbon; and impact parameter~($b$). The integration of the
Hamilton's equations of motion employed a time-step of 0.1\,\si{fs} such as to warrant
conservation of the total energy to better than \num{e-4}\,hartree\,(\si{\hartree}). Reactants were initially separated by 12\,\si{\bohr}, with a 
maximum value of $b$~($b_{max}$) optimized by trial and error for each $T$ 
and PES; see Tables~\ref{tab:isorates24}-\ref{tab:isorates}. 
Figure~\ref{fig:pess} shows sample reactive trajectories for  reactions~(\ref{eq:reac1})~and~(\ref{eq:reac3}). 

For a given $T$, (averaged) reaction cross-sections were then obtained as~\citep{PES99:171} 
\begin{equation}\label{eq:crossec}
\langle \sigma_{r}(T) \rangle = \pi b_{\rm max}^{2}\frac{N_{r}}{N}, 
\end{equation}
where $N_{r}$ is the number of reactive trajectories out of a total of $N$ that were run. To account in an 
approximate way for the deficiency of classical mechanics in 
conserving the quantum mechanical ZPE, we herein follow \citet{GUN90:2415} and \citet{VAR93:1076} and consider in the statistical analysis only trajectories that show enough vibrational energy to reach the ZPE of the products or the reformed reactants (Table~\ref{tab:diat}); no ZPE constraints were \emph{a priori} imposed on the $\mathrm{C_{3}}$ intermediate complexes (Table~\ref{tab:triat})~\citep{TRU79:188}. 

Assuming that the translational and internal degrees of freedom are at equilibrium, {that is,} the
velocity distributions are Maxwellian and the 
reactants quantum numbers are determined from Boltzmann distributions, the thermal rate coefficients of reactions~(\ref{eq:reac1})-(\ref{eq:reac4}) 
were calculated as~\citep{PES99:171}
\begin{equation}\label{eq:rate}
k(T)=g_{e}(T)\left(\frac{8k_{B}T}{\pi\mu_{\rm C+C_{2}}}\right)^{1/2}\langle \sigma_{r}(T) \rangle,
\end{equation}
with the estimated standard deviation (68.2\% error) given by
$\Delta k(T)\!=\!k(T)[(N\!-\!N_{r})/(NN_{r})]^{1/2}$. In Eq.~(\ref{eq:rate}), $k_{B}$ is the 
Boltzmann constant, $\mu_{\rm C+C_{2}}$ is the reactants
reduced mass and 
\begin{equation}\label{eq:gegen}
g_{e}(T)\!=\!\frac{Q_{e}(\mathrm{C_{3}})}{Q_{e}(\mathrm{C})\,Q_{e}(\mathrm{C_{2}})}
\end{equation}
is the electronic degeneracy factor {{that approximately accounts 
for fine structure effects~\citep{TRU72:3189,MUC72:3191,GRA90:2423,ZAN007:184308,ZAN010:9733}}}; the $Q_{e}$s are 
electronic partition functions. For $\mathrm{C_{3}}(^{1}A')$ and $\mathrm{C_{2}}(^{1}\Sigma^{+}_{g})$, they assume unit values, while $Q_{e}\left(\mathrm{C_{3}}(^{3}A')\right)\!=\!3$. For $\mathrm{C}(^{3}P)$ and $\mathrm{C_{2}}(^{3}\Pi_{u})$, the $Q_{e}$s are:  
\begin{equation}\label{eq:gec}
Q_{e}\left(\mathrm{C}(^{3}P)\right)=1+3\exp{\left(\frac{-23.62}{T}\right)}+5\exp{\left(\frac{-62.46}{T}\right)},
\end{equation}
and 
\begin{equation}\label{eq:gec2}
Q_{e}\left(\mathrm{C_{2}}(^{3}\Pi_{u})\right)=2+2\exp{\left(\frac{-21.97}{T}\right)}+2\exp{\left(\frac{-43.94}{T}\right)},
\end{equation}
where the first equation accounts for the populations of the $^{3}P_{J=0}$, $^{3}P_{J=1}$, and $^{3}P_{J=2}$ spin-orbit terms of $\mathrm{C}(^{3}P)$ with energy gaps 23.62 and 62.46\,\si{\kelvin} and degeneracy 2$J$+1~\citep{HAR017:16}. The corresponding inverted multiplets $^{3}\Pi_{\Omega\!=\!2}$, $^{3}\Pi_{\Omega\!=\!0}$, and $^{3}\Pi_{\Omega\!=\!1}$ of $\mathrm{C_{2}}(^{3}\Pi_{u})$ are considered in Eq.~(\ref{eq:gec2}); they are spaced by 21.97 
and 43.94\,\si{\kelvin} and are all doubly degenerate~\citep{BRO013:11,RAM014:5}. 
{{In deriving Eq.~(\ref{eq:gegen}), 
it is assumed that the spin-orbit states of the reactants are thermally
populated and that only specific fine-structure levels, that is, 
those that adiabatically correlate with the underlying PESs, may 
lead to reaction. For $\mathrm{C}(^{3}P)+\mathrm{C_{2}}(^{1}\Sigma^{+}_{g})$, 
we consider in Eq.~(\ref{eq:gegen}) that of the nine  
spin-orbit states arising asymptotically (Eq.~(\ref{eq:gec})) only the lowest three (correlating with the $^{3}A'$ PES) 
are reactive, these being the $\mathrm{C}(^{3}P_{0})+\mathrm{C_{2}}(^{1}\Sigma^{+}_{g})$ and two of the three $\mathrm{C}(^{3}P_{1})+\mathrm{C_{2}}(^{1}\Sigma^{+}_{g})$ states~\citep{WIL92:1886,RUS99:177}; 
for simplicity, no temperature dependence was \emph{a priori} included 
into the corresponding 
partition function,  that is, $Q_{e}\left(\mathrm{C_{3}}(^{3}A')\right)\!=\!3$ in Eq.~(\ref{eq:gegen})~\citep{WIL92:1886}.
The remaining six states 
correlate with two other 
excited triplet PESs and are regarded as nonreactive. 
Such a scenario becomes even more intricate in the case of 
$\mathrm{C}(^{3}P)+\mathrm{C_{2}}(^{3}\Pi_{u})$. Their asymptotic interaction gives rise to 18 (6 singlet, 6 triplet and 6 quintet) electronic states, correlating to a total of 54 spin-orbit levels (Eqs.~(\ref{eq:gegen})-(\ref{eq:gec2})). This undoubtedly makes the determination of the appropriate adiabatic correlations, and hence $Q_{e}\left(\mathrm{C_{3}}(^{1}A')\right)$ in Eq.~(\ref{eq:gegen}), 
a nontrivial task. 
Due to lack of experimental and further theoretical evidence, 
we herein simply choose to correlate the ground-state PES of C$_{3}$ to the lowest spin-obit states of its fragments~{\citep[\emph{i.e.}, to the lowest $A'$ component of $\mathrm{C}(^{3}P_{0})+\mathrm{C_{2}}(^{3}\Pi_{2})$;][]{AND003:5439,ABR008:4400}}, which means that this surface is the only one available for reaction among all 54 ($Q_{e}\left(\mathrm{C_{3}}(^{1}A')\right)\!=\!1$ in Eq.~(\ref{eq:gegen})). 
We note that while the above surmises are the most appealing \emph{a priori}, 
they may introduce, together with the single-surface ansatz~(\ref{eq:rate})~\citep{GRA90:2423}, additional approximations in the calculated rate coefficients; however, these can only be assessed once experimental kinetics data become available.
In this respect, we note that the possible contributions 
of the other excited states to the overall 
dynamics (not considered here) 
cannot be grasped at the moment as these and their associated global PESs remain largely unexplored. 
}}  
We further  note that we herein employ the same $Q_{e}$s for both main and rare isotopologs, which is a reasonable approximation. For example, the energy differences between spin-orbit terms of $\mathrm{^{12}C}(^{3}P)$ and $\mathrm{^{13}C}(^{3}P)$ and of $\mathrm{^{12}C_{2}}(^{3}\Pi_{u})$, $\mathrm{^{13}C^{12}C}(^{3}\Pi_{u}),$ and $\mathrm{^{13}C_{2}}(^{3}\Pi_{u})$ are well 
below 0.01\%~\citep{HAR017:16,BRO013:11,RAM014:5,AMI83:329}.

\section{Results and Discussion}\label{sec:results}
\begin{figure*}[htb!]
\captionsetup[subfigure]{labelformat=empty}
\centering
\subfloat{{\includegraphics[width=0.495\linewidth]{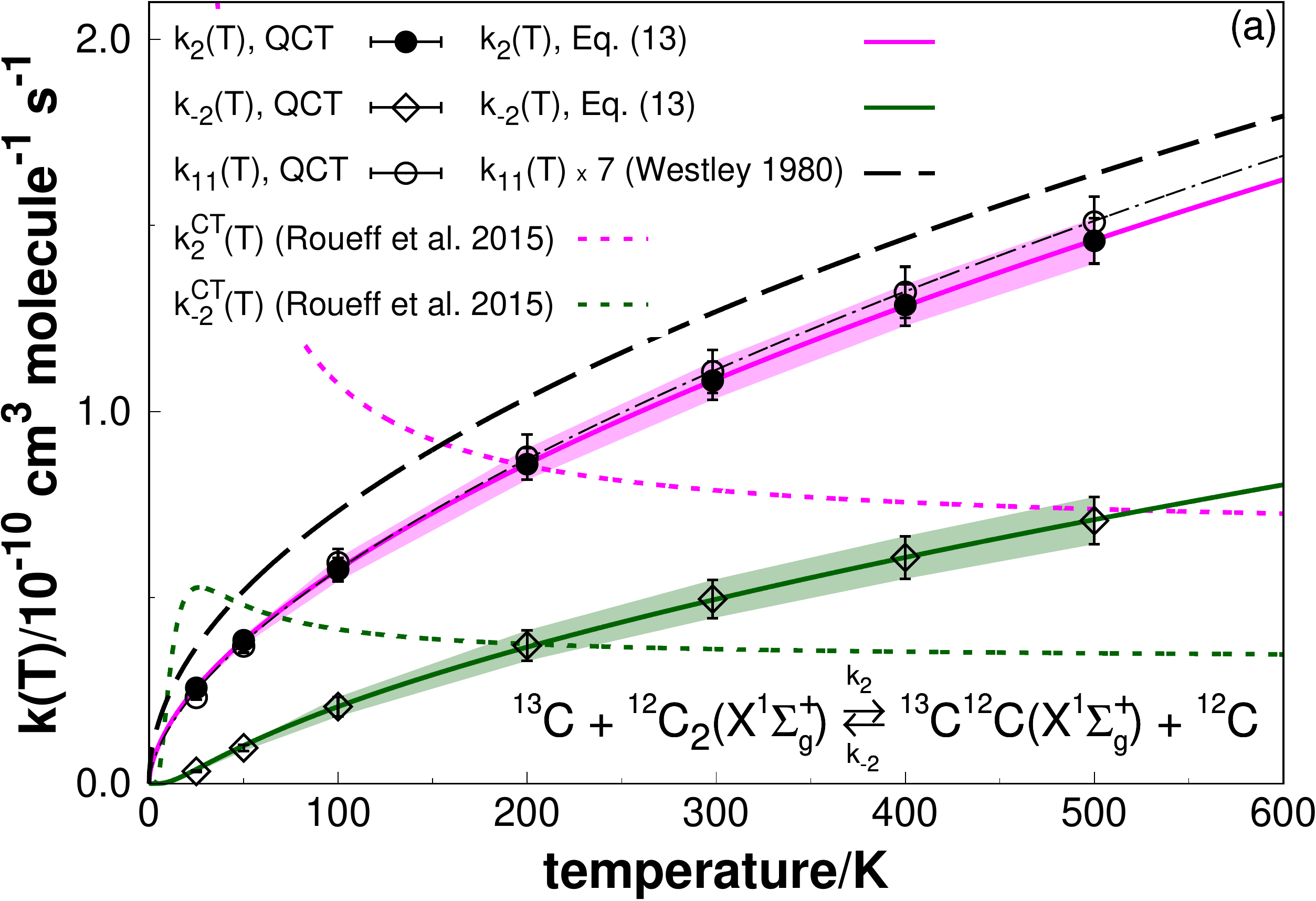}}}
\subfloat{{\includegraphics[width=0.495\linewidth]{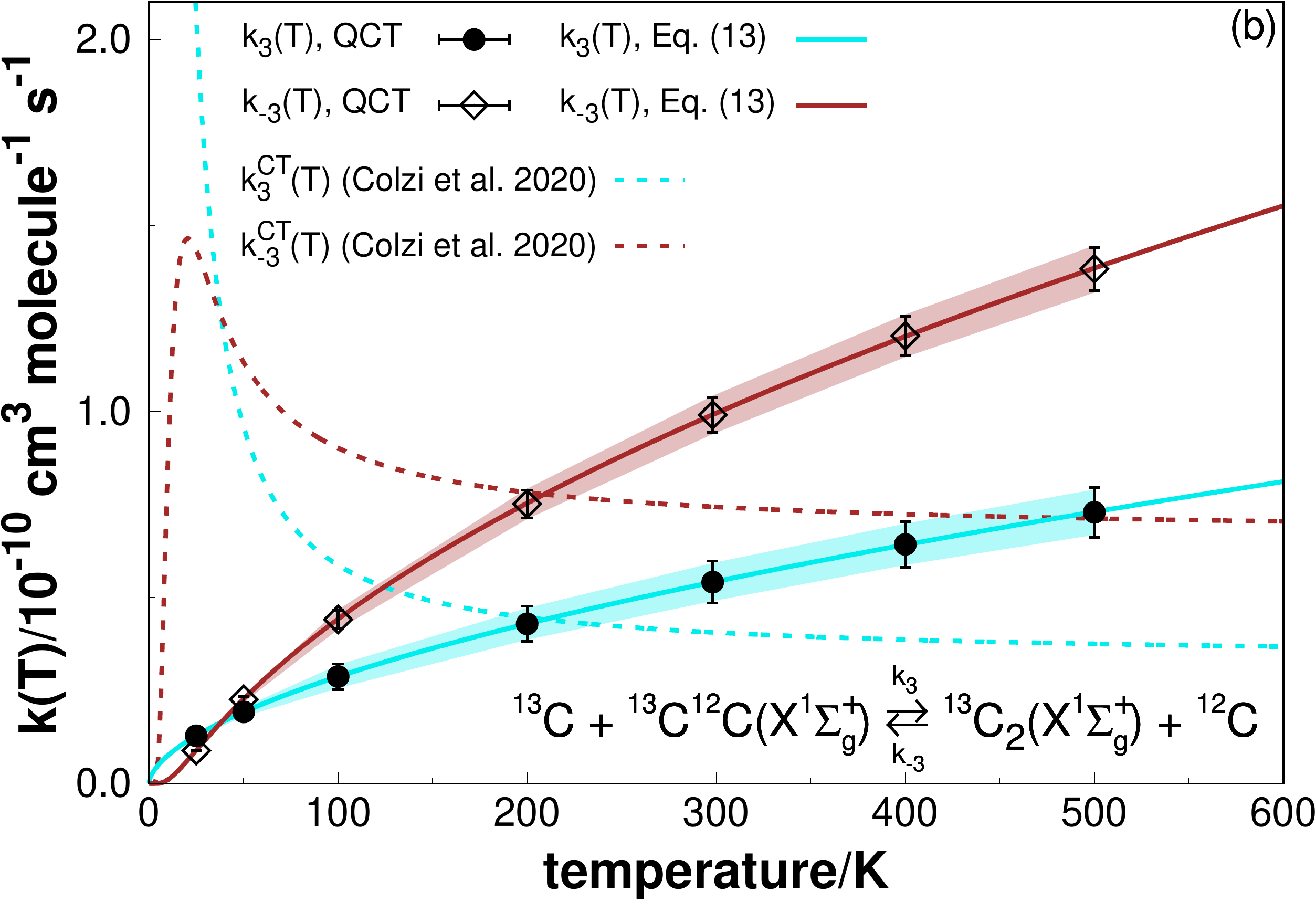}}}
\\
\subfloat{{\includegraphics[width=0.495\linewidth]{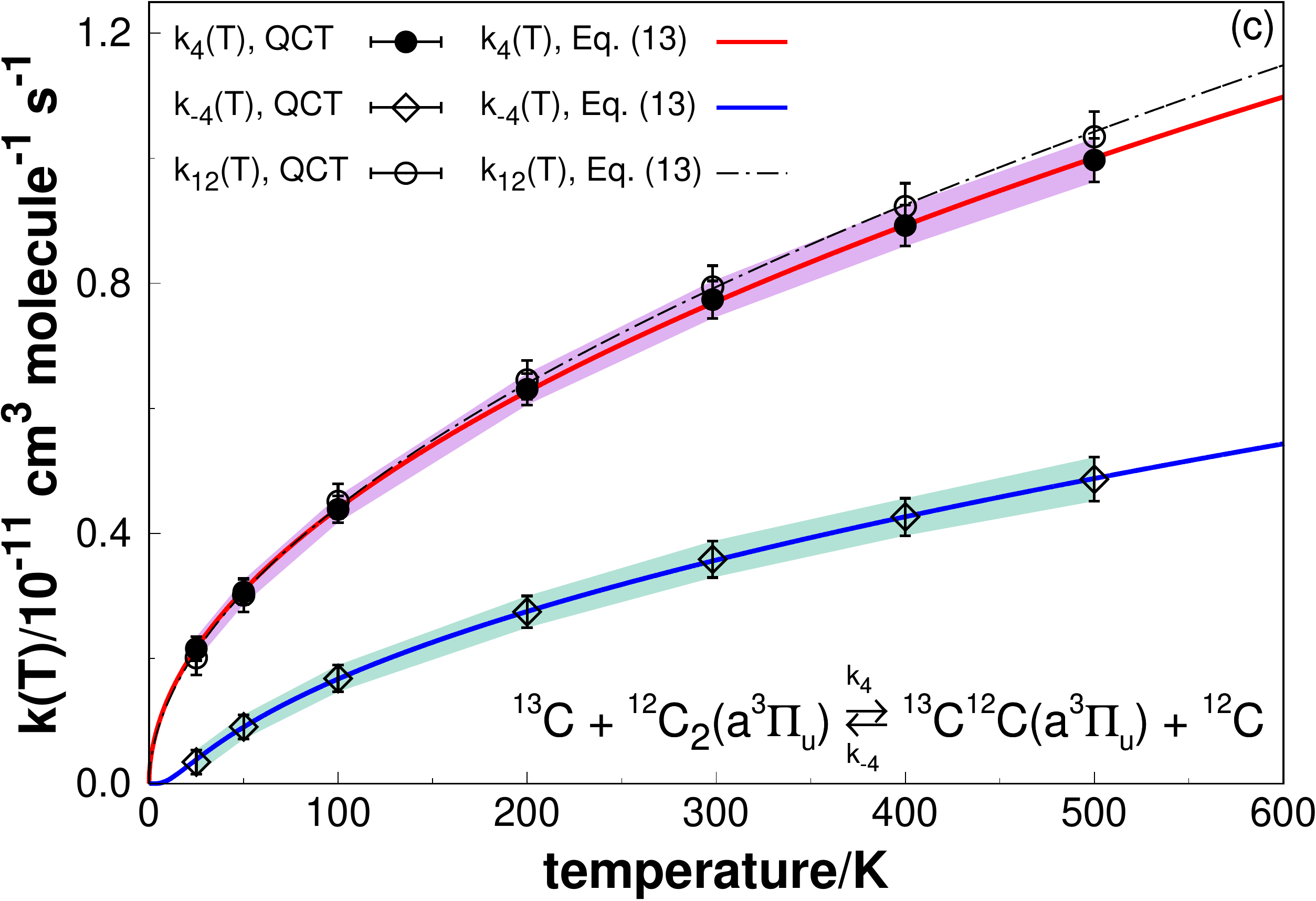}}}
\subfloat{{\includegraphics[width=0.495\linewidth]{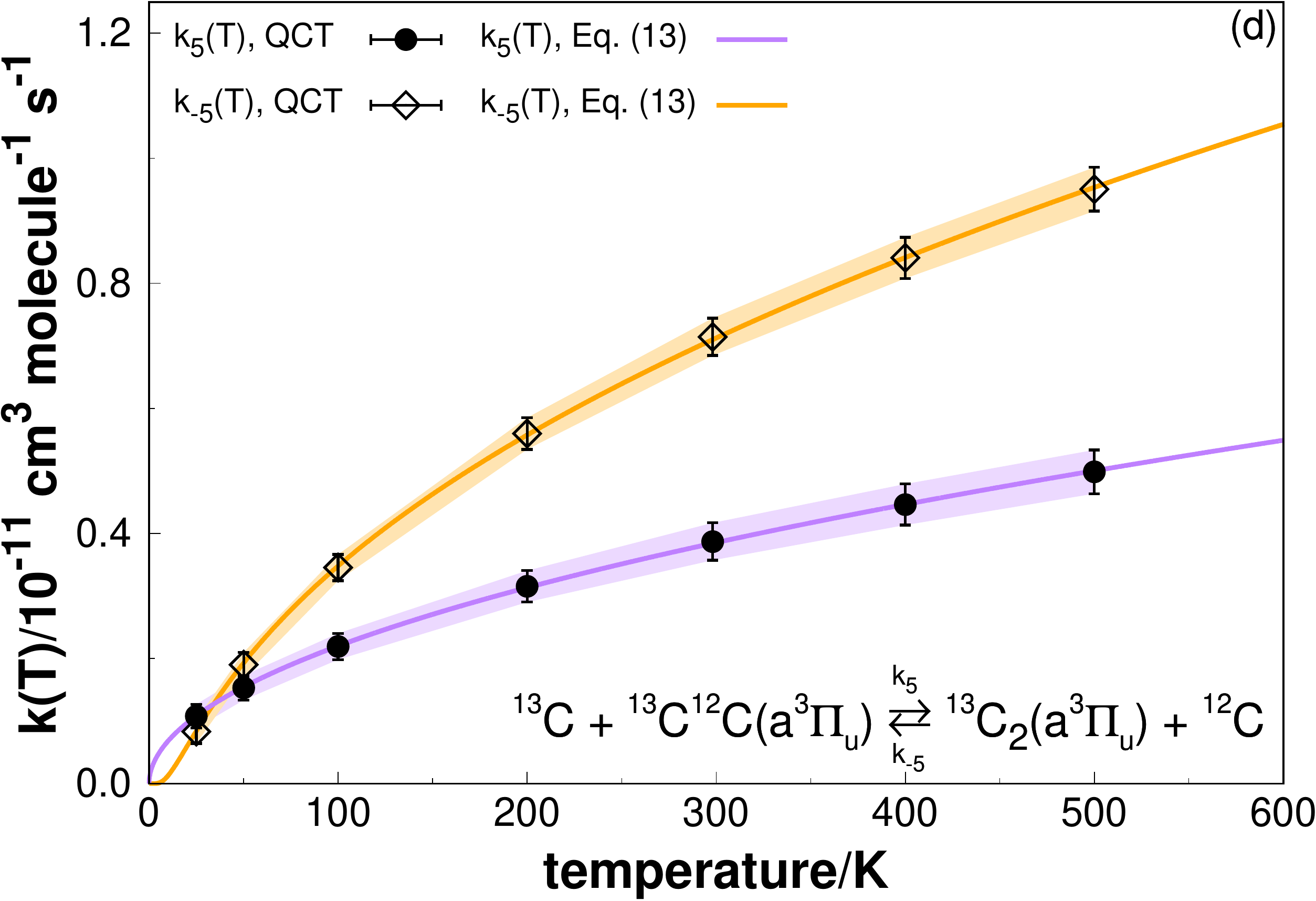}}}
\caption{\footnotesize Forward and backward thermal rate coefficients and associated error bars for the reactions~(a).~$\mathrm{^{13}C}(^{3}P)\!+\!\mathrm{^{12}C_{2}}(X^{1}\Sigma_{g}^{+}){\rightleftharpoons}\mathrm{^{13}C^{12}C}(X^{1}\Sigma_{g}^{+})\!+\!\mathrm{^{12}C}(^{3}P)$~[Eq.~(\ref{eq:reac1}), $k_{2,\text{-}2}$];~(b).~$\mathrm{^{13}C}(^{3}P)\!+\!\mathrm{^{13}C^{12}C}(X^{1}\Sigma_{g}^{+}){\rightleftharpoons}\mathrm{^{13}C_{2}}(X^{1}\Sigma_{g}^{+})\!+\!\mathrm{^{12}C}(^{3}P)$~[Eq.~(\ref{eq:reac2}), $k_{3,\text{-}3}$];~(c).~$\mathrm{^{13}C}(^{3}P)\!+\!\mathrm{^{12}C_{2}}(a^{3}\Pi_{u}){\rightleftharpoons}\mathrm{^{13}C^{12}C}(a^{3}\Pi_{u})\!+\!\mathrm{^{12}C}(^{3}P)$~[Eq.~(\ref{eq:reac3}), $k_{4,\text{-}4}$];~(d).~$\mathrm{^{13}C}(^{3}P)\!+\!\mathrm{^{13}C^{12}C}(a^{3}\Pi_{u}){\rightleftharpoons}\mathrm{^{13}C_{2}}(a^{3}\Pi_{u})\!+\!\mathrm{^{12}C}(^{3}P)$~[Eq.~(\ref{eq:reac4}), $k_{5,\text{-}5}$] at temperatures up to $600\,\si{\kelvin}$. 
Also shown are the QCT values obtained for the $\mathrm{^{12}C}+\mathrm{^{12}C_{2}}(X^{1}\Sigma_{g}^{+}/a^{3}\Pi_{u}){\rightarrow}\mathrm{^{12}C_{2}}(X^{1}\Sigma_{g}^{+}/a^{3}\Pi_{u})+\mathrm{^{12}C}$ atom-exchange reactions (Eqs.~(\ref{eq:atiso1}) and~(\ref{eq:atiso2})) and available results from the
literature~\citep{ROU015:A99,COL020,WES80:NSRDS}; CT stands for capture theory ~\citep{GEO005:194103}. Solid thick lines show the 
predicted QCT thermally averaged rates using the Arrhenius-Kooij 
formula of Eq.~(\ref{eq:rateforbac}).} 
\label{fig:rate}
\end{figure*}
Figure~\ref{fig:rate} shows the calculated 
forward and backward rate coefficients for the gas-phase isotope-exchange reactions~(\ref{eq:reac1})-(\ref{eq:reac4}) within the temperature range of $25\!\leq\!T/\si{\kelvin}\!\leq\!500$. Also shown for comparison are the corresponding QCT rates obtained for the
\begin{equation}\label{eq:atiso1}
\mathrm{^{12}C}+\mathrm{^{12}C_{2}}(X^{1}\Sigma_{g}^{+})\,{\stackrel{k_{11}}{\longrightarrow}}\,
\mathrm{^{12}C_{2}}(X^{1}\Sigma_{g}^{+})+\mathrm{^{12}C}, 
\end{equation}
and
\begin{equation}\label{eq:atiso2}
\mathrm{^{12}C}+\mathrm{^{12}C_{2}}(a^{3}\Pi_{u})\,{\stackrel{k_{12}}{\longrightarrow}}\,
\mathrm{^{12}C_{2}}(a^{3}\Pi_{u})+\mathrm{^{12}C}, 
\end{equation}
atom-exchange reactions and available results from the
literature~\citep{ROU015:A99,COL020,WES80:NSRDS}; Tables~\ref{tab:isorates24}-\ref{tab:isorates} gather all the  
numerical values. To further explore the temperature dependence of $k$, we have considered the popular 
Arrhenius-Kooij formula~\citep{LAI84:494}
\begin{equation}\label{eq:rateforbac}
k(T)=A\left(\frac{T}{298.15}\right)^{B}\exp{\left(\frac{-C}{T}\right)},
\end{equation}
where $A$, $B$, and $C$ are parameters to be 
adjusted to the QCT 
data; they are numerically defined in
Table~\ref{tab:rateparam}, with the final fitted forms also plotted in Figure~\ref{fig:rate}.
 We note that, in the least-squares fitting procedure, the nonlinear parameters $C$ were allowed to float freely from their initial values, and 
therefore slightly deviate from the expected 
$\Delta E_{\mathrm{ZPE}}$ values in Table~\ref{tab:zpe}. Physically, 
this is consistent with the presence of rotationally excited 
reactant and product $\mathrm{C_{2}}$ species~\citep{MLA014:A144}. 
Suffice it to say that, due to the homonuclear 
nature of the $\mathrm{^{12/13}C_{2}(X^{1}\Sigma_{g}^{+})}$ reactant molecules, 
only even rotational quantum numbers $J$ were 
considered in the trajectory samplings; 
{for the $\mathrm{^{13}C^{12}C}$ species}, the corresponding Boltzmann distributions include both odd and even $J$ values.
\begin{table}
\centering
\caption{\footnotesize Parameters of Eq.~(\ref{eq:rateforbac}) for the forward and reverse rate coefficients of reactions~(\ref{eq:reac1})-(\ref{eq:reac4}),~(\ref{eq:atiso1}), and~(\ref{eq:atiso2}).}
\label{tab:rateparam}
\begin{threeparttable}
\begin{tabular}{c
S[table-align-text-post=false,table-format=2.8]
S[table-align-text-post=false,table-format=2.8]
S[table-align-text-post=false,table-format=2.8]}
\hline\hline
 Rate     &  \multicolumn{3}{c}{Parameter\tnote{a}} \\
                       \cline{2-4}\\[-0.4cm]
constant & {$A$} & {$B$} & {$C$} \\ 
\hline\\[-0.4cm]
$k_{\ref{eq:reac1}}$         &  1.0824\,{($-10$)} & 5.7905\,{($-1$)} &  0\\  
$k_{\text{-}\ref{eq:reac1}}$ &  5.3988\,{($-11$)} & 6.3165\,{($-1$)} &   2.6963\,{($+1$)}\\
$k_{\ref{eq:reac2}}$         &  5.4118\,{($-11$)} & 5.7905\,{($-1$)} &  0\\
$k_{\text{-}\ref{eq:reac2}}$ &  1.0835\,{($-10$)} & 5.7742\,{($-1$)} &   2.6560\,{($+1$)}\\
$k_{\ref{eq:reac3}}$         &  7.6852\,{($-12$)} & 5.1035\,{($-1$)} &  0\\
$k_{\text{-}\ref{eq:reac3}}$ &  3.8553\,{($-12$)} & 5.4722\,{($-1$)} &  2.3706\,{($+1$)}\\
$k_{\ref{eq:reac4}}$         &  3.8426\,{($-12$)} & 5.1035\,{($-1$)} &  0\\
$k_{\text{-}\ref{eq:reac4}}$ &  7.6919\,{($-12$)} & 5.0770\,{($-1$)} &  2.3850\,{($+1$)}\\
$k_{\ref{eq:atiso1}}$        &  1.1078\,{($-10$)} & 6.0149\,{($-1$)} &  0\\
$k_{\ref{eq:atiso2}}$        &  7.9168\,{($-12$)} & 5.3256\,{($-1$)} &  0\\
\hline\hline
\end{tabular}
\begin{tablenotes}[flushleft]
  \item[a]{{\footnotesize $x\,(y)$ represents $x\times10^{y}$. $A$ is in $\rm cm^{3}\,molecule^{-1}\,s^{-1}$, $B$ unitless and $C$ is in $\si{\kelvin}$}.} 
\end{tablenotes}
\end{threeparttable}
\end{table}
\begin{figure*}
\captionsetup[subfigure]{labelformat=empty}
\centering
\subfloat{{\includegraphics[width=0.495\linewidth]{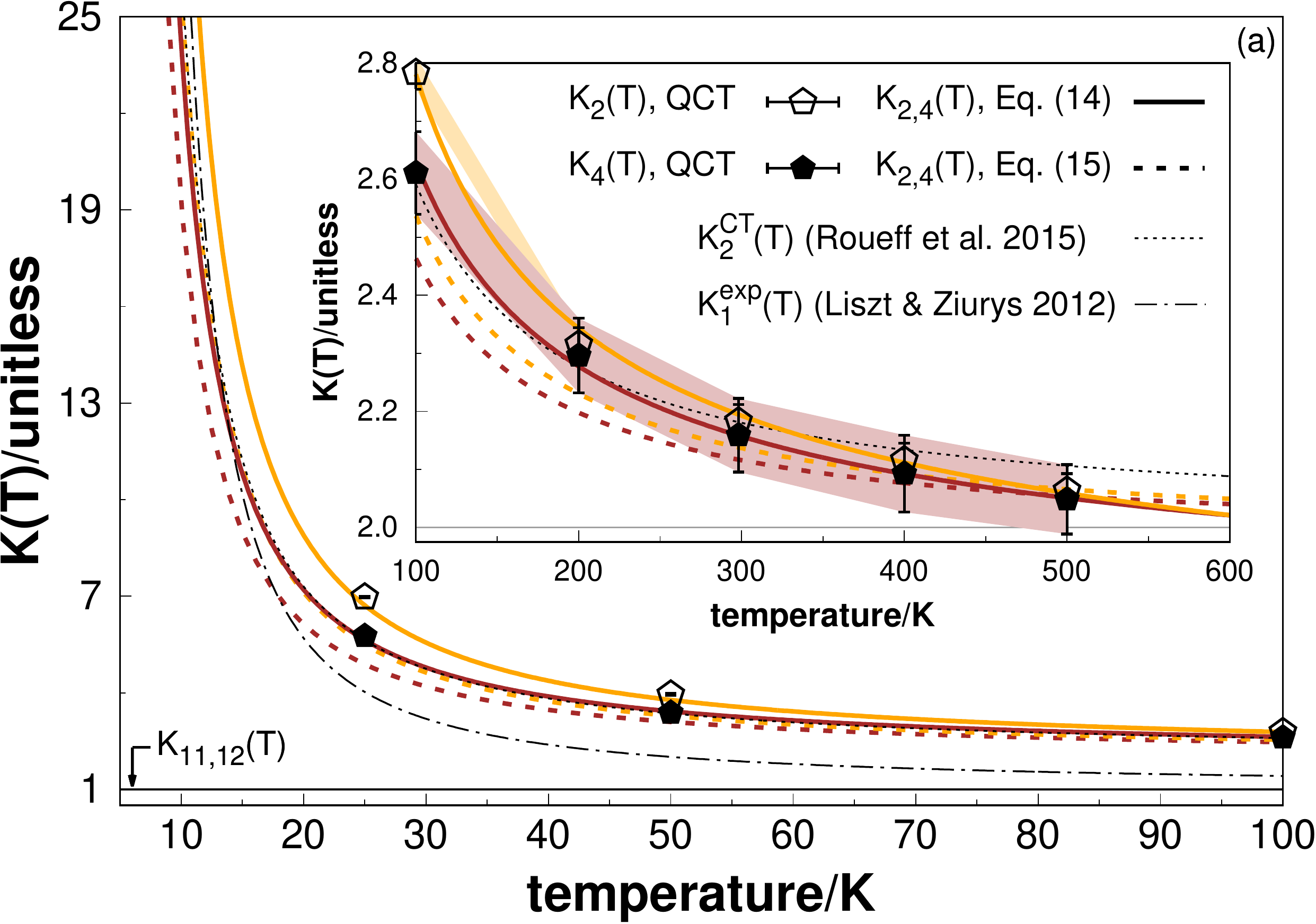}}}
\hfill
\subfloat{{\includegraphics[width=0.495\linewidth]{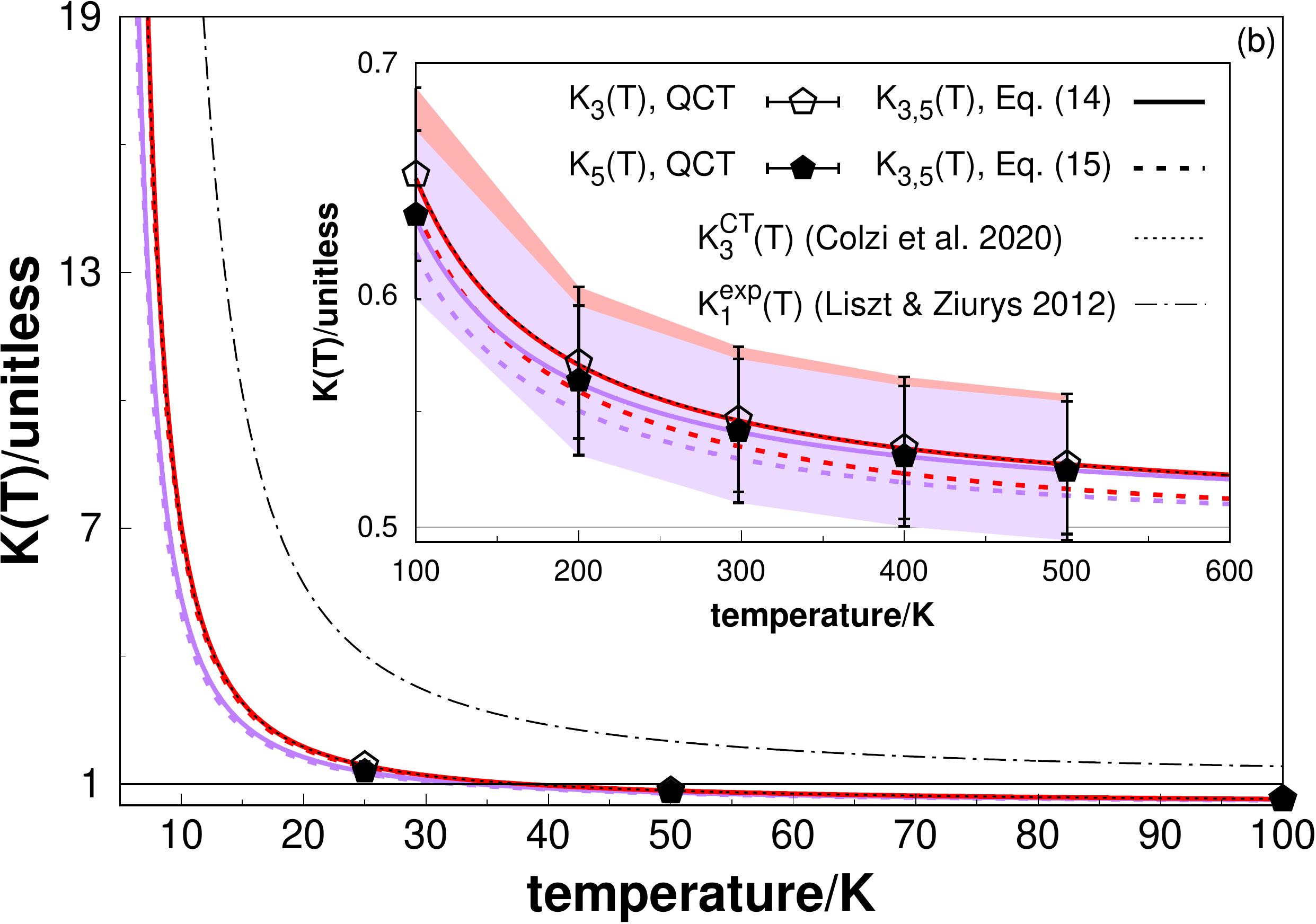}}}
\caption{\footnotesize Equilibrium constants and associated error bars 
for the reactions~(a).~$\mathrm{^{13}C}(^{3}P)\!+\!\mathrm{^{12}C_{2}}(X^{1}\Sigma_{g}^{+}){\rightleftharpoons}\mathrm{^{13}C^{12}C}(X^{1}\Sigma_{g}^{+})\!+\!\mathrm{^{12}C}(^{3}P)$~[Eq.~(\ref{eq:reac1}), $K_{2}\!=\!k_{2}/k_{\text{-}2}$]~and~$\mathrm{^{13}C}(^{3}P)\!+\!\mathrm{^{12}C_{2}}(a^{3}\Pi_{u}){\rightleftharpoons}\mathrm{^{13}C^{12}C}(a^{3}\Pi_{u})\!+\!\mathrm{^{12}C}(^{3}P)$~[Eq.~(\ref{eq:reac3}), $K_{4}\!=\!k_{4}/k_{\text{-}4}$];~(b).~$\mathrm{^{13}C}(^{3}P)\!+\!\mathrm{^{13}C^{12}C}(X^{1}\Sigma_{g}^{+}){\rightleftharpoons}\mathrm{^{13}C_{2}}(X^{1}\Sigma_{g}^{+})\!+\!\mathrm{^{12}C}(^{3}P)$~[Eq.~(\ref{eq:reac2}), $K_{3}\!=\!k_{3}/k_{\text{-}3}$]~and~$\mathrm{^{13}C}(^{3}P)\!+\!\mathrm{^{13}C^{12}C}(a^{3}\Pi_{u}){\rightleftharpoons}\mathrm{^{13}C_{2}}(a^{3}\Pi_{u})\!+\!\mathrm{^{12}C}(^{3}P)$~[Eq.~(\ref{eq:reac4}), $K_{5}\!=\!k_{5}/k_{\text{-}5}$] at temperatures up to $600\,\si{\kelvin}$. Points and solid thick lines are obtained from Eq.~(\ref{eq:eqconst}) using the QCT thermally averaged rates and their analytic forms in Eq.~(\ref{eq:rateforbac}), respectively, while dashed thick lines represent theoretical estimates based on statistical mechanics (Eq.~(\ref{eq:eqconstest})). Also shown are the corresponding values 
predicted via capture theory~\citep{ROU015:A99,COL020} for reactions~(\ref{eq:reac1})~and~(\ref{eq:reac2})   
and experimental data (exp) for $\mathrm{^{13}C^{+}}+\mathrm{^{12}CO}{\rightleftharpoons}\mathrm{^{13}CO}+\mathrm{^{12}C^{+}}$ (Eq.~(\ref{eq:reacbase}), $K_{1}\!=\!k_{1}/k_{\text{-}1}$) as taken from \citet{LIS012:55}. For clarity, the $K$ values for the thermoneutral reactions~(\ref{eq:atiso1}) and~(\ref{eq:atiso2}) are also indicated. High-temperature limits are represented by gray solid lines.} 
\label{fig:eqconst}
\end{figure*}

As shown in Figure~\ref{fig:rate}, the calculated 
thermal rate constants for the $\mathrm{C}\!+\!\mathrm{C_{2}}$ reactions increase as a function of temperature, revealing a positive $T$ dependence. As 
previously noted (section~\ref{subsec:pess}), this stems 
from the fact that, at higher $T$, not only are the (head-on collinear) MEPs sampled by the reactive trajectories 
but also other regions of the PESs become energetically accessible (\emph{e.g.}, bimolecular side-on encounters at high collision energies), increasing reaction probabilities. 
A similar temperature-dependent profile ($k\!\propto\!T^{0.6}$) was found experimentally for the barrierless $\mathrm{N}\!+\!\mathrm{C_{2}}$ reaction~\citep{LOI014:14212}. As expected, all these processes evolve via long-lived 
trajectories, with the strongly bound energized 
complexes spanning large sections of the molecular PESs; see Figure~\ref{fig:pess}. Figure~\ref{fig:rate} shows that 
the forward 
exothermic reactions~(\ref{eq:reac1})-(\ref{eq:reac4}) are fast with the calculated rate constants varying 
from $10^{-12}$ up to $10^{-10}\,\rm cm^{3}\,molecule^{-1}\,s^{-1}$ within the temperature interval considered.  
At $T\!=\!10\,\si{\kelvin}$, Eq.~(\ref{eq:rateforbac}) predicts $k_{2}$, $k_{3}$, $k_{4}$, $k_{5}$ to be 
$\num{1.5e-11}$, $\num{7.6e-12}$, $\num{1.4e-12}$, and $\num{6.8e-13}\,\rm cm^{3}\,molecule^{-1}\,s^{-1}$, respectively; these values are typical of atom--radical reactions that are currently included in low-temperature astrochemical networks~\citep{SMI004:323}. 
Overall, the reactivity of $\mathrm{C_{2}}(X^{1}\Sigma_{g}^{+})$ with ground-state C atoms is about one order of magnitude higher than that of the first excited $\mathrm{C_{2}}(a^{3}\Pi_{u})$ state. This is in general agreement with experimental 
results when the molecular partner is an unsaturated hydrocarbon~\citep{GU006:245,PAR008:9591}. We further note that, except for $T\!=\!200\,\si{\kelvin}$ (see Figure~\ref{fig:rate}~(a~and~b)), the predicted rates of the isotope-exchange reactions~(\ref{eq:reac1})~and~(\ref{eq:reac2}) are in sharp contrast to 
the theoretically derived $k$ via simple capture theory~{\citep[CT;][]{ROU015:A99,COL020}}, particularly at low $T$. Such discrepancies are large enough to suggest that, in addition to long-range interactions, the strongly bound (short-range) parts  
of the PESs considered  here also influence the dynamics of all these reactive processes. One should bear in mind that, although 
an approximate treatment of the 
ZPE-leakage~\citep{TRU79:188} is warranted here (see section~\ref{subsec:qct}), our QCT approach (like CT~\citep{GEO005:194103}) neglects, by its own nature, other quantum-mechanical (QM) effects such as tunneling; this is also justifiable on the large masses of the nuclei  
involved. While such an approximation may be less reliable in   
the low-temperature limit~\citep{TRU79:505,PES99:171}, 
accurate estimates of QM effects unavoidably
require exact (nonadiabatic) quantum dynamics 
calculations which are even more demanding in the case of  
complex-forming reactions~\citep{GUO012:1}, and hence are beyond {the present} scope of this work.

In contrast to the 
forward reactions, the backward processes in Eqs.~(\ref{eq:reac1})-(\ref{eq:reac4})
show temperature thresholds (Table~\ref{tab:rateparam});  
these latter are 
attributed to ZPE 
differences between reactant and product $\mathrm{C_{2}}$ 
isotopologs. 
Due to operation of 
statistical factors on the kinetics of~(\ref{eq:reac1})~and~(\ref{eq:reac3})   
(\emph{i.e.}, $\frac{1}{2}$ for backward and 
$1$ for forward), we recognize from Figure~\ref{fig:rate}~(a~and~c) that, in 
the high-$T$ limit, 
the rate coefficients $k_{\text{-}2,\text{-}4}$ are approximately half of $k_{2,4}$~\citep{HEN81:1201}. The contrary is the case for  reactions~(\ref{eq:reac2})~and~(\ref{eq:reac4}) where 
statistical factors of $1$ for backward and 
$\frac{1}{2}$ for the forward processes are operative~\citep{HEN81:1201}. Therefore, as shown in Figure~\ref{fig:rate}~(b~and~d), $k_{\text{-}3,\text{-}5}\!\approx\!2k_{3,5}$ in the high-$T$ limit.  
However, at lower temperatures,  the manifestation of the statistical factors on all these rate coefficients is largely 
masked 
by the increased influence 
of such $T$ thresholds~\citep{HEN81:1201}.

The (small) effects of the isotope substitution on the overall 
kinetics (\emph{i.e.}, the kinetic-isotope effect) 
can primarily be assessed from Figure~\ref{fig:rate}~(a~and~c). 
By comparing the thermoneutral reactions~(\ref{eq:atiso1}) and~(\ref{eq:atiso2}) with  
the forward ones in Eqs.~(\ref{eq:reac1})~and~(\ref{eq:reac3}), one can 
see that, given the lower ZPE content of the 
$\mathrm{^{13}C^{12}C}$ product species and the exothermic nature of these latter pair of reactions, abstraction by $\mathrm{^{13}C}(^{3}P)$ is slightly faster than by $\mathrm{^{12}C}(^{3}P)$ at low $T$. Nevertheless, such an energy defect ($\Delta E_{\mathrm{ZPE}}$)  becomes 
less significant in determining reactivity as long as higher internal and 
collision energies are accessible at higher $T$.  
We note that the calculated thermal rate coefficients of reaction~(\ref{eq:atiso1}) are about seven times greater than those reported by \citet{WES80:NSRDS}. 

To quantify the possible impact of reactions~(\ref{eq:reac1})-(\ref{eq:reac4}) on
the overall C fractionation chemistry, in Figure~\ref{fig:eqconst} we plot their equilibrium constants ($K$) as a function of the temperature. These were obtained using  
both QCT data and the analytic forms in Eq.~(\ref{eq:rateforbac}) as  
\begin{equation}\label{eq:eqconst}
K(T)=\frac{k_{f}(T)}{k_{r}(T)}\equiv\frac{[\mathrm{^{12}C}][\mathrm{^{P}C_{2}}]}{[\mathrm{^{13}C}][\mathrm{^{R}C_{2}}]},
\end{equation}
where $k_{f}$ and $k_{r}$ are the forward and reverse rates, with  
R and P identifying the corresponding reactant and product $\mathrm{C_{2}}$ isotopolog. These $K$ values are also compared with theoretical estimates based on statistical mechanics~\citep{TER000:563,MLA014:A144,MLA017:A22},
\begin{equation}\label{eq:eqconstest}
K(T)=f_{m}^{3/2}\frac{Q_{\text{int}}(\mathrm{^{P}C_{2}})}{Q_{\text{int}}(\mathrm{^{R}C_{2}})}\exp{\left(\frac{\Delta E_{\mathrm{ZPE}}}{T}\right)},
\end{equation}
where the mass factor $f_{m}$ is given by
\begin{equation}\label{eq:massfac}
f_{m}=\frac{m(\mathrm{^{12}C})m(\mathrm{^{P}C_{2}})}{m(\mathrm{^{13}C})m(\mathrm{^{R}C_{2}})}, 
\end{equation}
with $m(X)$ denoting the mass of the species $X$; $\Delta E_{\mathrm{ZPE}}$ in Eq.~(\ref{eq:eqconstest}) is in \si{\kelvin}. The internal partition
function, $Q_{\text{int}}$, includes only the rovibrational degrees 
of freedom (no translation and electronic contributions) and 
is given by the standard expression,
\begin{equation}\label{eq:partfunc}
Q_{\text{int}}=g_{\Lambda,\text{hfs}}\sum_{v}\sum_{J}(2J+1)\mathrm{e}^{-\epsilon^{J}_{v}/k_{B}T}, 
\end{equation}
where $\epsilon^{J}_{v}$ is the diatomic rovibrational energy (with total angular momentum $J$ and vibrational quantum number $v$) measured relative to the corresponding ZPE; this is calculated from the experimentally derived two-body term of the associated $\mathrm{C_{3}}$ PES. In Eq.~(\ref{eq:partfunc}), $g_{\Lambda,\text{hfs}}$ accounts for the combined effects of 
$\Lambda$-doubling and nuclear spin (hyperfine) degeneracy and is  
defined in~{\citet[][see Table~3 therein]{IRW87:348}}.  
For comparison, we also plot in Figure~\ref{fig:eqconst} equilibrium constants for reactions~(\ref{eq:reac1})~and~(\ref{eq:reac2})   
obtained via CT~\citep{ROU015:A99,COL020}   
and the experimental values of $\mathrm{^{13}C^{+}}+\mathrm{^{12}CO}{\rightleftharpoons}\mathrm{^{13}CO}+\mathrm{^{12}C^{+}}$ taken from~\citet{LIS012:55}. 

The data presented in Figure~\ref{fig:eqconst} 
clearly indicate that the C isotopic fractionation 
occurs most efficiently at low 
temperatures, notably in reactions~(\ref{eq:reac1})~and~(\ref{eq:reac3}). Under these conditions virtually   
all the available $\mathrm{^{13}C}$ is in the form of 
$\mathrm{^{13}C^{12}C}$, with only a small fraction being 
locked up in $\mathrm{^{13}C_{2}}$. 
Among $\mathrm{^{13}C^{12}C}$, ground-state $\mathrm{^{13}C^{12}C}(^{1}\Sigma_{g}^{+})$ appears to be the dominant species owing to the higher exothermicity of reaction~(\ref{eq:reac1}); see Table~\ref{tab:rateparam}. Indeed, by extrapolating Eq.~(\ref{eq:rateforbac}) in~(\ref{eq:eqconst}) to the typical temperature of dense clouds, $T\!=\!10\,\si{\kelvin}$, we obtain $K_{2}\!\approx\!36$, $K_{4}\!\approx\!24$, $K_{3}\!\approx\!7$, and $K_{5}\!\approx\!5$. These former values are quite close to the one predicted for the ion--molecule $\mathrm{^{13}C^{+}}+\mathrm{^{12}CO}$
reaction (Eq.~(\ref{eq:reacbase})), $K_{1}\!\approx\!33$~\citep{LAN84:581}. We note that, in the high-$T$ limit, the equilibrium constants converge to well-defined values: 2 for the isotope-exchange reactions~(\ref{eq:reac1})~and~(\ref{eq:reac3}) and $\frac{1}{2}$ for (\ref{eq:reac2})~and~(\ref{eq:reac4}). Such limits reflect 
the manifestation of the aforementioned statistical factors in the overall chemical kinetics and become equivalent to `symmetry' (or probability) factors appearing in previous statistical thermodynamic considerations~\citep{TER000:563}. In this regard, we note that the calculated $K$values  from Eq.~(\ref{eq:eqconstest}) represent lower limits to the actual QCT data and are roughly consistent (as expected) with the ones predicted from CT~\citep{ROU015:A99,COL020}. We reiterate that, similarly to Eq.~(\ref{eq:eqconstest}), CT does not take into account all the details of the molecular PESs in estimating the macroscopic kinetic and thermodynamic attributes. 

\section{Astrophysical implications}\label{sec:molabund}
\begin{figure}
\centering
\includegraphics[angle=0,width=1\linewidth]{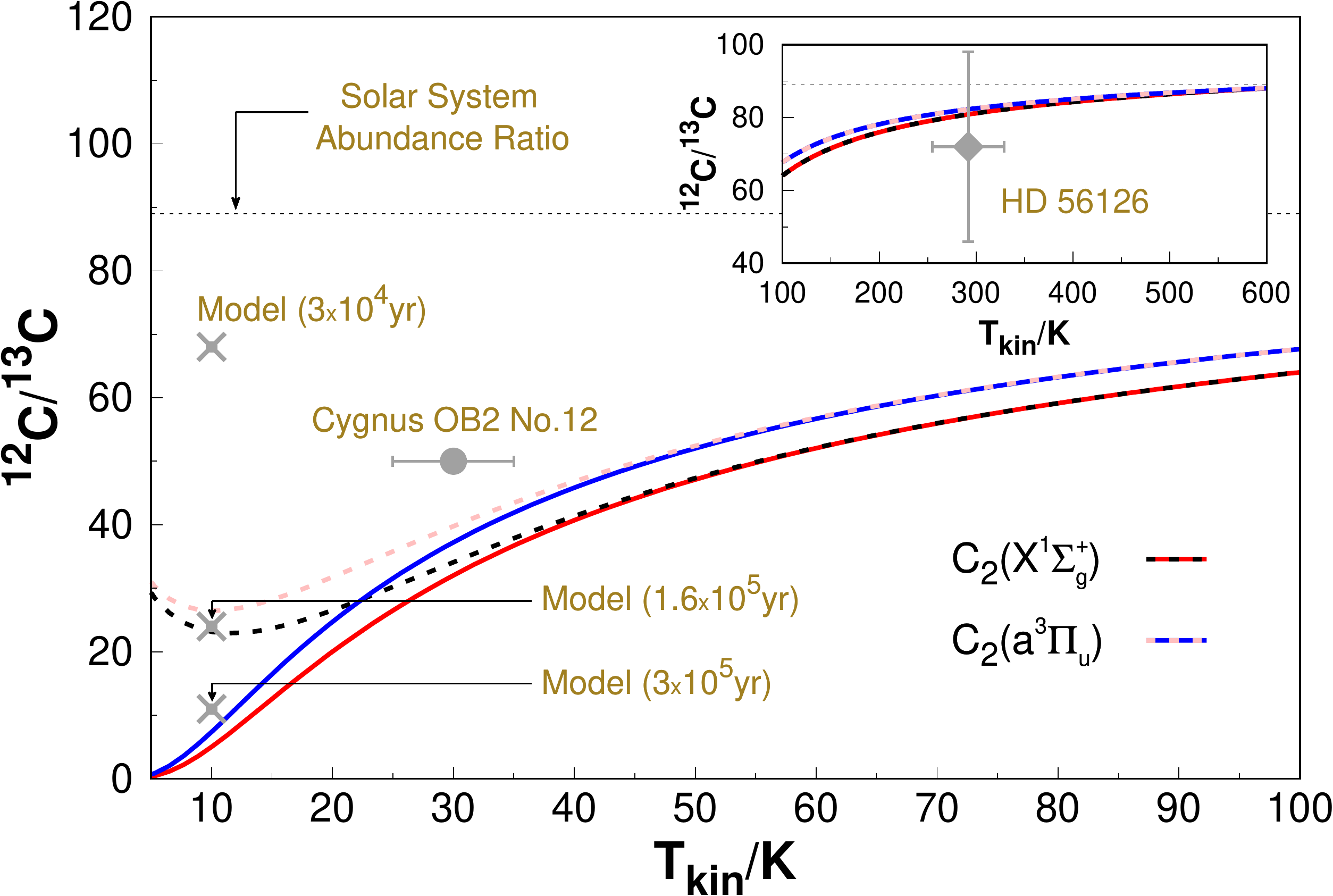}
\caption{\footnotesize Variation of the $^{12}$C/$^{13}$C isotope ratios derived from $\mathrm{C_{2}}(X^{1}\Sigma_{g}^{+})$ and 
$\mathrm{C_{2}}(a^{3}\Pi_{u})$ (by means of reactions~(\ref{eq:reac1})   
and~(\ref{eq:reac3}), respectively) as a function of the 
gas kinetic temperature ($T_{\rm kin}$). Solid (blue and red) lines represent the theoretical values calculated from Eq.~(\ref{eq:cratio}) assuming chemical equilibrium conditions, while the corresponding (pink and back) dashed lines show their behavior as obtained from a reduced kinetic model with fixed integration time of $\num{1.6e5}$\,yr (see text). Also shown by the gray points (with error bars) are the corresponding values obtained from observational surveys conducted by~\citet{HAM019:143} and~\citet{BAK98:387} towards Cyg~OB2~No.~12 and HD~56126, respectively, as well as 
those reported by~\citet{COL020} for $10\,\si{\kelvin}$ using a gas-grain chemical model in three different simulation timescales. The horizontal 
dashed line (also in the inset) highlights the elemental [$^{12}$C/$^{13}$C] Solar System abundance ratio.  
} 
\label{fig:ratio}
\end{figure}
To further (qualitatively) assess  the extent to which  
the most relevant reactions~(\ref{eq:reac1})   
and~(\ref{eq:reac3}) influence the net $^{13}$C chemical enrichment in diverse astronomical environments and their possible effects on observational data, we plot in Figure~\ref{fig:ratio} the expected theoretical 
$^{12}$C/$^{13}$C atomic carbon ratios versus kinetic temperature ($T_{\rm kin}$) as possibly measured from $\mathrm{C_{2}}$ (\emph{e.g.}, via its Phillips ($A\,^{1}\Pi_{u}$--$X\,^{1}\Sigma_{g}^{+}$) and Swan ($d\,^{3}\Pi_{g}$--$a\,^{3}\Pi_{u}$) bands).  
Following~\citet{SMI80:424}, the calculated ratios were obtained from Eq.~(\ref{eq:eqconst}), that is, assuming chemical equilibrium conditions 
\begin{equation}\label{eq:cratio}
\frac{^{12}\mathrm{C}}{^{13}\mathrm{C}}(T)=2\times \frac{[\mathrm{^{12}C}]}{[\mathrm{^{13}C}]}\times\frac{1}{K(T)}\equiv 2\times \frac{[^{12}\mathrm{C}_{2}]}{[^{13}\mathrm{C}^{12}\mathrm{C}]}(T),
\end{equation}
where $K(T)$ are the corresponding equilibrium 
constants ($K_{2}$ and $K_{4}$ for the $X\,^{1}\Sigma_{g}^{+}$ and $a\,^{3}\Pi_{u}$ states, respectively, see, \emph{e.g.}, Figure~\ref{fig:eqconst}~(a)) and 
$[\mathrm{^{12}C}]/[\mathrm{^{13}C}]$ is the elemental (reservoir) carbon abundance ratio taken to be equal to the Solar System value of 89; the factor of 2 appears due to statistical considerations; see, \emph{e.g.},~\citet{BAK98:387}. For comparison, we also show 
the corresponding values obtained from observational surveys on $\mathrm{C_{2}}$ isotopologs conducted by~\citet{HAM019:143} in the context of 
translucent clouds (\emph{i.e.}, in the line of sight of Cyg~OB2~No.~12) and~\citet{BAK98:387} towards the circumstellar envelope of the post-AGB star 
HD~56126. As emphasized by~\citet{HAM019:143}, their work reports the first marginal detection of $^{13}\mathrm{C}^{12}\mathrm{C}$ in the ISM. Due to the lack of 
observational data on $[^{12}\mathrm{C}_{2}]/[^{13}\mathrm{C}^{12}\mathrm{C}]$ in molecular clouds, we resort to the $^{12}$C/$^{13}$C ratios derived from $\mathrm{C_{2}}(X^{1}\Sigma_{g}^{+})$ by~\citet{COL020} using a time-dependent gas-grain chemical model; {the model results} are also plotted in Figure~\ref{fig:ratio} 
for three different simulation timescales. 
Figure~\ref{fig:ratio} shows that, although the calculated $^{12}$C-to-$^{13}$C ratios depict slightly varying degrees of fractionation depending on whether they are inherited from $\mathrm{C_{2}}(X^{1}\Sigma_{g}^{+})$ or $\mathrm{C_{2}}(a^{3}\Pi_{u})$, the general profiles are both consistent with a $^{13}$C-enhancement at the lower temperatures of interstellar clouds. However, we note that, at even lower $T_{\rm kin}$, all $^{12}$C/$^{13}$C ratios drop to very small values; this is not necessary true in reality given that interstellar chemistry may unavoidably deviate from thermodynamic equilibrium. {To gauge   
the impact of such a departure from equilibrium on the calculated ratios, 
we follow~\citet{SMI80:424} and impose time dependence on $^{12}$C/$^{13}$C by integrating analytically the corresponding kinetic differential (continuity) equations for $\mathrm{^{13}C^{12}C}(X^{1}\Sigma_{g}^{+})$ [Eq.~(\ref{eq:reac1})] and $\mathrm{^{13}C^{12}C}(a^{3}\Pi_{u})$ [Eq.~(\ref{eq:reac3})]; for brevity, 
the final formulas are not be given here, and we refer the reader to Eqs.~(12) and~(13) of~\citet{SMI80:424} for details. 
The theoretical $^{12}$C/$^{13}$C ratios obtained in this way 
are shown by the dashed lines in Figure~\ref{fig:ratio}. We note that in solving the corresponding rate equations, we assume $[\mathrm{^{12}C}]/[\mathrm{^{13}C}]$ as terrestrial (as in Eq.~(\ref{eq:cratio})) and   
consider a fixed integration time of $\num{1.6e5}$\,yr with a $\mathrm{^{12}C}$ fractional abundance of $\num{1e-5}$~;~these latter parameters  
are both consistent with an early cloud chemistry~\citep{COL020}.  
On the basis of these assumptions, Figure~\ref{fig:ratio} reveals a clear mismatch between the calculated early chemistry and equilibrium $^{12}$C-to-$^{13}$C ratios for $T_{\rm kin}\!\lessapprox\!30\,\si{\kelvin}$. However, for larger temperatures,  reactive equilibrium appears to be promptly reached; see Figure~\ref{fig:ratio}. Moreover, the plotted data from~\citet{COL020} indicate that the predicted ratios from chemical kinetics also converge (as expected) to those at equilibrium for longer simulation times. 
Yet, at $10\,\si{\kelvin}$, our theoretical $^{12}$C/$^{13}$C ratio derived from $\mathrm{C_{2}}(X^{1}\Sigma_{g}^{+})$ agrees quite well   
with the value reported by~\citet{COL020} within the $\num{1.6e5}$\,yr timescale. 
}
As for the 
observational data, the calculated $^{12}$C/$^{13}$C ratios show fairly good 
correlations with those given by~\citet{BAK98:387} and~\citet{HAM019:143}. The larger deviations observed towards Cyg~OB2~No.~12 (see Figure~\ref{fig:ratio}) provide further evidence that, besides $^{13}$C+$\mathrm{C_{2}}$ chemical fractionation, other competing photo-induced processes and/or secondary reactions are at work in translucent clouds; reportedly, one should also take into account the large uncertainties in the measurements by~\citet{HAM019:143}. As highlighted by these latter authors, future observations of $\mathrm{^{12}C^{13}C}$ using higher quality spectra will provide a 
clear picture on the $\mathrm{C_{2}}$ carbon isotope ratios in the ISM. 
Meanwhile, the determination of accurate laboratory and theoretical 
reaction rate coefficients for the most efficient fractionation pathways like $^{13}$C+$\mathrm{C_{2}}$ and $^{13}$C+$\mathrm{C_{3}}$~\citep{GIE020:A120,COL020} would be useful for the interpretation of interstellar C fractionation chemistry via astrochemical models~\citep{ROU015:A99,COL020,LOI020}.

\section{Summary}\label{sec:conclusions}
In the present work, we provide accurate theoretical rate  
coefficients as a function of the temperature for all possible isotope-exchange reactions of C with $\mathrm{C_{2}}(X^{1}\Sigma_{g}^{+},a^{3}\Pi_{u})$. To this end, we used the quasi-classical trajectory method, with the previously obtained (mass-independent) PESs of $\mathrm{C_{3}}(^{3}A',^{1}A')$ providing the required forces between the colliding partners. The calculated 
rate coefficients within the range of $25\!\leq\!T/\si{\kelvin}\!\leq\!500$ exhibit a positive temperature dependence and {our results show a behavior that clearly differs from  previous theoretical estimates based on simple capture theory~\citep{ROU015:A99,COL020}. This suggests that, in addition to long-range interactions, the strongly bound (short-range) parts  
of the underlying PESs also influence the dynamics of the reactive processes.} For each reaction considered, analytic three-parameter Arrhenius-Kooij formulas are derived that readily interpolate and extrapolate the associated forward and reverse rates. 
To quantify their possible impact on the interstellar C isotopic chemistry, equilibrium constants of all such processes are evaluated 
from the calculated kinetics data, unraveling their increased efficiency 
into $^{13}$C incorporation at low $T$. {For the most relevant reactions and assuming both equilibrium and time-dependent conditions, theoretical $^{12}$C/$^{13}$C atomic  carbon ratios as a function of the gas kinetic temperature are also reported and compared with available model chemistry and observational data on $\mathrm{C_{2}}$.}  
Despite some previous claims~\citep{BAK98:387}, the present theoretical 
results strongly support the suggestion made by other authors~\citep{ROU015:A99,COL020} that 
the $\mathrm{C}+\mathrm{C_{2}}$ reactions (particularly~(\ref{eq:reac1})   
and~(\ref{eq:reac3})) may act as important routes in the overall C-fractionation chemistry, notably in low-temperature C-rich environments. 
Besides providing key input data for astrochemical models of cold dense clouds~\citep{FUR011:38,ROU015:A99,COL020,LOI020}, 
the calculated rate constants over such a broad $T$ range may also fulfill  
the needs of models of photo-dissociation regions~\citep{ROL013:A56}, translucent clouds~\citep{HAM019:143}, protoplanetary disks~\citep{WOO009:1360}, and circumstellar envelopes of evolved C-stars~\citep{BAK98:387}. Apart from 
its astrophysical implications, this work is expected to  
provide safe grounds on which to base future methodological developments toward the calculation of theoretical rate constants of astrochemically relevant isotope-exchange reactions without resorting to (and avoid the burden of) quantum dynamics, while still recovering all intrinsic details of the interacting potentials between the colliding particles.

\begin{acknowledgements}
   This work has received funding from the {European Union's Horizon 2020 research and innovation program under the Marie Sklodowska-Curie} grant agreement No 894321.
\end{acknowledgements}

%
%


\onecolumn
\begin{appendix}
\section{Tables}
\begin{table}[htb!]
\centering
\caption{\footnotesize Maximum impact parameters, and
thermal rate and equilibrium 
constants as a function of the temperature of  
the isotope-exchange reactions~(\ref{eq:reac1})~and~(\ref{eq:reac3}).}
\label{tab:isorates24}
\begin{threeparttable}
\begin{tabular}{
S[table-align-text-post=false,table-format=3.2]
S[table-align-text-post=false,table-format=1.1]
S[table-align-text-post=false,table-format=2.9]
S[table-align-text-post=false,table-format=2.9]
S[table-align-text-post=false,table-format=1.4]}
\hline\hline
\\
\multicolumn{5}{c}{$\mathrm{^{13}C}(^{3}P)+\mathrm{^{12}C_{2}}(X^{1}\Sigma_{g}^{+})\underset{k_{\text{-}2}}{\stackrel{k_{2}}{\rightleftharpoons}}\mathrm{^{13}C^{12}C}(X^{1}\Sigma_{g}^{+})+\mathrm{^{12}C}(^{3}P)+\Delta E_{\mathrm{ZPE}}(=\!25.8\,\si{\kelvin})$} \\
 \\
 \hline
{$T/\si{\kelvin}$} & {$b_{max}$/$\si{\angstrom}$} & {$k_{2}$/$\rm cm^{3}\,molecule^{-1}\,s^{-1}$} & {$k_{\text{-}2}$/$\rm cm^{3}\,molecule^{-1}\,s^{-1}$} & {$K_{2}$/unitless} \\ 
\hline
500.00 & 5.0 & 1.4581\,{($-10$)\tnote{a}} & 7.0643\,{($-11$)} & 2.0641 \\
400.00 & 5.0 & 1.2857\,{($-10$)}          & 6.0748\,{($-11$)} & 2.1164 \\
298.15 & 5.1 & 1.0826\,{($-10$)}          & 4.9588\,{($-11$)} & 2.1831 \\
200.00 & 5.0 & 0.8584\,{($-10$)}          & 3.7070\,{($-11$)} & 2.3156 \\
100.00 & 5.0 & 0.5750\,{($-10$)}          & 2.0655\,{($-11$)} & 2.7842 \\
 50.00 & 4.9 & 0.3850\,{($-10$)}          & 0.9623\,{($-11$)} & 4.0008 \\
 25.00 & 4.6 & 0.2570\,{($-10$)}          & 0.3299\,{($-11$)} & 7.7902 \\ 
\hline  
\\
\multicolumn{5}{c}{$\mathrm{^{13}C}(^{3}P)+\mathrm{^{12}C_{2}}(a^{3}\Pi_{u})\underset{k_{\text{-}4}}{\stackrel{k_{4}}{\rightleftharpoons}}\mathrm{^{13}C^{12}C}(a^{3}\Pi_{u})+\mathrm{^{12}C}(^{3}P)+\Delta E_{\mathrm{ZPE}}(=\!22.9\,\si{\kelvin})$} \\
 \\
 \hline
{$T/\si{\kelvin}$} & {$b_{max}$/$\si{\angstrom}$} & {$k_{4}$/$\rm cm^{3}\,molecule^{-1}\,s^{-1}$} & {$k_{\text{-}4}$/$\rm cm^{3}\,molecule^{-1}\,s^{-1}$} & {$K_{4}$/unitless} \\ 
\hline\\[-0.4cm]
500.00 & 5.0 & 9.9709\,{($-12$)} & 4.8677\,{($-12$)} & 2.0484 \\
400.00 & 5.0 & 8.9250\,{($-12$)} & 4.2646\,{($-12$)} & 2.0928 \\
298.15 & 5.0 & 7.7400\,{($-12$)} & 3.5854\,{($-12$)} & 2.1588 \\
200.00 & 5.0 & 6.3078\,{($-12$)} & 2.7473\,{($-12$)} & 2.2960 \\
100.00 & 5.0 & 4.3868\,{($-12$)} & 1.6802\,{($-12$)} & 2.6109 \\
 50.00 & 4.9 & 3.0643\,{($-12$)} & 0.9066\,{($-12$)} & 3.3799 \\
 25.00 & 4.8 & 2.1578\,{($-12$)} & 0.3448\,{($-12$)} & 6.2581 \\
\hline\hline
\end{tabular}
\begin{tablenotes}[flushleft]
  \item[a]{{\footnotesize $x\,(-y)$ represents $x\times10^{-y}$}.} 
\end{tablenotes}
\end{threeparttable}
\end{table}
\begin{table}[htb!]
\centering
\caption{\footnotesize Maximum impact parameters, and
thermal rate and equilibrium 
constants as a function of the temperature of  
the isotope-exchange reactions~(\ref{eq:reac2})~and~(\ref{eq:reac4}).}
\label{tab:isorates35}
\begin{threeparttable}
\begin{tabular}{
S[table-align-text-post=false,table-format=3.2]
S[table-align-text-post=false,table-format=1.1]
S[table-align-text-post=false,table-format=2.9]
S[table-align-text-post=false,table-format=2.9]
S[table-align-text-post=false,table-format=1.4]}
\hline\hline
\\
\multicolumn{5}{c}{$\mathrm{^{13}C}(^{3}P)+\mathrm{^{13}C^{12}C}(X^{1}\Sigma_{g}^{+})\underset{k_{\text{-}3}}{\stackrel{k_{3}}{\rightleftharpoons}}\mathrm{^{13}C_{2}}(X^{1}\Sigma_{g}^{+})+\mathrm{^{12}C}(^{3}P)+\Delta E_{\mathrm{ZPE}}(=\!26.3\,\si{\kelvin})$} \\
 \\
 \hline
{$T/\si{\kelvin}$} & {$b_{max}$/$\si{\angstrom}$} & {$k_{3}$/$\rm cm^{3}\,molecule^{-1}\,s^{-1}$} & {$k_{\text{-}3}$/$\rm cm^{3}\,molecule^{-1}\,s^{-1}$} & {$K_{3}$/unitless} \\ 
\hline
500.00 & 5.3 & 7.2907\,{($-11$)\tnote{a}} & 1.3827\,{($-10$)} & 0.5273 \\
400.00 & 5.2 & 6.4284\,{($-11$)}          & 1.2031\,{($-10$)} & 0.5343 \\
298.15 & 5.4 & 5.4131\,{($-11$)}          & 0.9903\,{($-10$)} & 0.5466 \\
200.00 & 5.1 & 4.2920\,{($-11$)}          & 0.7516\,{($-10$)} & 0.5710 \\
100.00 & 5.3 & 2.8747\,{($-11$)}          & 0.4408\,{($-10$)} & 0.6521 \\
 50.00 & 5.1 & 1.9250\,{($-11$)}          & 0.2263\,{($-10$)} & 0.8505 \\
 25.00 & 4.9 & 1.2850\,{($-11$)}          & 0.0888\,{($-10$)} & 1.4471 \\ 
\hline  
\\
\multicolumn{5}{c}{$\mathrm{^{13}C}(^{3}P)+\mathrm{^{13}C^{12}C}(a^{3}\Pi_{u})\underset{k_{\text{-}5}}{\stackrel{k_{5}}{\rightleftharpoons}}\mathrm{^{13}C_{2}}(a^{3}\Pi_{u})+\mathrm{^{12}C}(^{3}P)+\Delta E_{\mathrm{ZPE}}(=\!23.5\,\si{\kelvin})$} \\
 \\
 \hline
{$T/\si{\kelvin}$} & {$b_{max}$/$\si{\angstrom}$} & {$k_{5}$/$\rm cm^{3}\,molecule^{-1}\,s^{-1}$} & {$k_{\text{-}5}$/$\rm cm^{3}\,molecule^{-1}\,s^{-1}$} & {$K_{5}$/unitless} \\ 
\hline\\[-0.4cm]
500.00 & 5.2 & 4.9854\,{($-12$)} & 9.5064\,{($-12$)} & 0.5244 \\
400.00 & 5.2 & 4.4625\,{($-12$)} & 8.4084\,{($-12$)} & 0.5307 \\
298.15 & 5.2 & 3.8700\,{($-12$)} & 7.1450\,{($-12$)} & 0.5416 \\
200.00 & 5.1 & 3.1539\,{($-12$)} & 5.5987\,{($-12$)} & 0.5633 \\
100.00 & 5.2 & 2.1934\,{($-12$)} & 3.4560\,{($-12$)} & 0.6347 \\
 50.00 & 5.0 & 1.5321\,{($-12$)} & 1.9018\,{($-12$)} & 0.8056 \\
 25.00 & 4.9 & 1.0789\,{($-12$)} & 0.8312\,{($-12$)} & 1.2980 \\
\hline\hline
\end{tabular}
\begin{tablenotes}[flushleft]
  \item[a]{{\footnotesize $x\,(-y)$ represents $x\times10^{-y}$}.} 
\end{tablenotes}
\end{threeparttable}
\end{table}
\begin{table}[htb!]
\centering
\caption{\footnotesize Maximum impact parameters and 
thermal rate 
constants as a function of the temperature for 
the atom-exchange reactions~(\ref{eq:atiso1})~and~(\ref{eq:atiso2}).}
\label{tab:isorates}
\begin{threeparttable}
\begin{tabular}{
S[table-align-text-post=false,table-format=3.2]
S[table-align-text-post=false,table-format=1.1]
S[table-align-text-post=false,table-format=2.9]
}
\hline\hline
\\
\multicolumn{3}{c}{$\mathrm{^{12}C}(^{3}P)+\mathrm{^{12}C_{2}}(X^{1}\Sigma_{g}^{+})\,{\stackrel{k_{11}}{\longrightarrow}}\,\mathrm{^{12}C_{2}}(X^{1}\Sigma_{g}^{+})+\mathrm{^{12}C}(^{3}P)$} \\
 \\
 \hline
{$T/\si{\kelvin}$} & {$b_{max}$/$\si{\angstrom}$} & {$k_{11}$/$\rm cm^{3}\,molecule^{-1}\,s^{-1}$} \\ 
\hline
500.00 & 4.9 & 1.5087\,{($-10$)\tnote{a}} \\
400.00 & 4.9 & 1.3202\,{($-10$)}          \\
298.15 & 5.0 & 1.1071\,{($-10$)}          \\
200.00 & 4.9 & 0.8774\,{($-10$)}          \\
100.00 & 4.9 & 0.5935\,{($-10$)}          \\
 50.00 & 4.8 & 0.3693\,{($-10$)}          \\
 25.00 & 4.7 & 0.2312\,{($-10$)}          \\ 
\hline  
\\
\multicolumn{3}{c}{$\mathrm{^{12}C}(^{3}P)+\mathrm{^{12}C_{2}}(a^{3}\Pi_{u})\,{\stackrel{k_{12}}{\longrightarrow}}\,\mathrm{^{12}C_{2}}(a^{3}\Pi_{u})+\mathrm{^{12}C}(^{3}P)$} \\
 \\
 \hline
{$T/\si{\kelvin}$} & {$b_{max}$/$\si{\angstrom}$} & {$k_{12}$/$\rm cm^{3}\,molecule^{-1}\,s^{-1}$} \\ 
\hline
500.00 & 4.9 &  1.0344\,{($-11$)} \\
400.00 & 4.9 &  0.9228\,{($-11$)} \\
298.15 & 4.9 &  0.7942\,{($-11$)} \\
200.00 & 4.9 &  0.6456\,{($-11$)} \\
100.00 & 5.0 &  0.4514\,{($-11$)} \\
 50.00 & 4.8 &  0.3015\,{($-11$)} \\
 25.00 & 4.7 &  0.2012\,{($-11$)} \\
\hline\hline
\end{tabular}
\begin{tablenotes}[flushleft]
  \item[a]{{\footnotesize $x\,(-y)$ represents $x\times10^{-y}$}.} 
\end{tablenotes}
\end{threeparttable}
\end{table}
\end{appendix}

\end{document}